\documentclass[lettersize,journal]{IEEEtran}
\usepackage{amsmath,amsfonts}
\usepackage{array}
\usepackage[caption=false,font=normalsize,labelfont=sf,textfont=sf]{subfig}
\usepackage{textcomp}
\usepackage{stfloats}
\usepackage{url}
\usepackage{verbatim}
\usepackage{graphicx}
\usepackage{cite}
\usepackage[ruled,linesnumbered]{algorithm2e}
\usepackage{tikz}
\usepackage{mathtools}
\usepackage{makecell}
\usepackage{empheq}
\usepackage{booktabs}
\usepackage{multirow}
\usepackage{threeparttable}
\usepackage{multicol,lipsum}
\usepackage{colortbl}

\newcommand{\tian}[1]{\textcolor{black}{#1}}
\newcommand{\tianNew}[1]{\textcolor{black}{#1}}

\newcommand{\model}[1]{\textit{SmartSplit}}
\newcommand*{\circled}[1]{\lower.7ex\hbox{\tikz\draw (0pt, 0pt)%
    circle (.5em) node {\makebox[1em][c]{\small #1}};}}

\begin{document}

\title{Breaking the Memory Wall for Heterogeneous Federated Learning via Model Splitting}

\author{Chunlin Tian, Li Li, Kahou Tam, Yebo Wu, Cheng-Zhong Xu, Fellow, IEEE}

\markboth{Journal of \LaTeX\ Class Files,~Vol.~14, No.~8, August~2021}%
{Shell \MakeLowercase{\textit{et al.}}: A Sample Article Using IEEEtran.cls for IEEE Journals}

\maketitle

\begin{abstract}
Federated Learning (FL) enables multiple devices to collaboratively train a shared model while preserving data privacy. Ever-increasing model complexity coupled with limited memory resources on the participating devices severely bottlenecks the deployment of FL in real-world scenarios. Thus, a framework that can effectively break the memory wall while jointly taking into account the hardware and statistical heterogeneity in FL is urgently required.

In this paper, we propose \model~, a framework that effectively reduces the memory footprint on the device side while guaranteeing the training progress and model accuracy for heterogeneous FL through model splitting.
Towards this end, \model~ employs a hierarchical structure to adaptively guide the overall training process. In each training round, the central manager, hosted on the server, dynamically selects the participating devices and sets the cutting layer by jointly considering the memory budget, training capacity, and data distribution of each device. The MEC manager, deployed within the edge server, proceeds to split the local model and perform training of the server-side portion. Meanwhile, it fine-tunes the splitting points based on the time-evolving statistical importance. The on-device manager, embedded inside each mobile device, continuously monitors the local training status while employing cost-aware checkpointing to match the runtime dynamic memory budget. \tian{Extensive experiments on representative datasets are conducted on both commercial off-the-shelf mobile device testbeds. The experimental results show that \model~ excels in FL training on highly memory-constrained mobile SoCs, offering up to a 94\% peak latency reduction and 100-fold memory savings. It enhances accuracy performance by 1.49\%-57.18\% and adaptively adjusts to dynamic memory budgets through cost-aware recomputation} 
\end{abstract}


\section{introduction}
Federated Learning (FL)~\cite{2016federated} shines the spotlight on a new machine learning paradigm that efficiently coordinates multiple mobile devices (e.g., smartphones and wearable devices) to train a collaborative DNN model while preserving data privacy. \tian{A lot of works~\cite{model_size,Middleware} have suggested that DNNs with more complex architectures and larger sizes can effectively improve the quality of analysis.} Thus, \tian{recently developed DNNs are becoming deeper and wider.} Meanwhile, larger memory space is required to store the parameters, intermediate outputs, and gradients during the training process~\cite{Low-memory}. For instance, training ResNet152 requires 5.58 GB with a batch size of 32. However, the development of mobile systems does not keep at the same pace. The available RAM for existing mobile devices is quite limited, only ranging from 4 to 16 GB~\cite{RAM}. Ever-increasing model sizes coupled with limited memory resources, unfortunately, exclude clients with low-end devices that would otherwise make their contribution to the shared model with their local data. \tian{Moreover, the overall training process cannot even be triggered for current mainstream models due to high memory requirements, such as those needed for transformers~\cite{Transformer}.}

\textbf{Limitation of Existing Approaches.} In order to reduce the memory footprint during the training procedure, several memory optimization techniques have been proposed which can be broadly divided into the following categories: 1) gradient checkpointing~\cite{checking_1,checking_2}, 2) micro-batching~\cite{Sage, accumulate}, 3) model size reduction~\cite{mixed,int_2} and 4) host-device memory virtualization~\cite{vitrual_1,vitrual_2}. Gradient checkpointing releases a subset of intermediate activations and recomputes non-stored activations on demand during backpropagation. However, recomputation introduces extra computation latency. Micro-batching breaks up large batch sizes to reduce activation footprint, however utilizing parallelism for memory results in costly bandwidth consumption that affects training efficiency. Model size reduction lessens or scales down parameters to save the model, optimizer, and activation footprint, however, lossy training affects model performance by sacrificing accuracy and affecting convergence. Host-device memory virtualization expands the memory budget with host-side memory, however, mobile SoCs adopt a unified memory scheme in which the CPU and GPU share the same physical memory and contend during concurrent data access. \tian{Therefore, while existing approaches can significantly reduce the memory footprint, they severely compromise training efficiency or model accuracy, rendering them unsuitable for supporting an FL system.} \textit{Thus, a framework that can jointly take into account memory reduction, training efficiency, and model accuracy is critical for FL in real-world scenarios.}

\textbf{Observations and Challenges.} We observe that model splitting can be a feasible approach to effectively reduce the memory footprint. It splits the model into two parts. The front part is kept on the device side in order to preserve data privacy while the latter part can be offloaded to the server. The activation output of the cut layer and the gradients are exchanged between the mobile device and the central server during the training procedure. However, effectively applying model splitting in a highly dynamic and heterogeneous FL training environment still faces the following critical challenges. First, the data distributions and hardware configurations are totally heterogeneous across the devices. Moreover, different layers in the collaborative model usually require totally unique computing and memory resources to complete the training process. Thus, how to select the participating devices and the cutting layer for each device to optimize the training efficiency and model accuracy in a unified manner is the first challenge. 
Second, model splitting introduces high communication overhead between the model device and the central server during each training iteration which severely slows down the training progress. Thus, effectively reducing the communication overhead while guaranteeing the model's accuracy is also challenging. Lastly, resource contention from the concurrently running apps sets a highly dynamic memory budget for the training of the device-side portion. \tian{Ensuring that the training process can proceed successfully under these constraints is another critical challenge.}

In this paper, we propose~\model, a novel hierarchical management system for memory-friendly FL on mobile devices via model splitting. It mainly consists of the following three components. (1) \textit{Central Manager} hosted on the server: it selects participants and dynamically splits the training model based on fine-grained estimates of runtime system profiles, statistical availability, and within memory thresholds. (2) \textit{Mobile Edge Computing (MEC) Manager} inside the edge server: it mitigates the communication overhead of the split smashed layer and explores the evolution of data importance to enhance the FL training efficiency. (3) \textit{On-device Manager} deployed on each mobile device: it monitors system performance on-the-fly and executes scheduled training as specified by the advanced managers while employing cost-aware recomputation for run-time memory management to improve the local memory consumption. \tian{
Extensive experiments on representative datasets are conducted to evaluate \model~ utilizing both commercial off-the-shelf hardware and simulation testbeds. Compared with the baselines, \model~ enables FL training over extremely memory-constrained mobile SoCs.  It boasts a staggering 94\% peak latency reduction, hastening model convergence, and provides up to a 100-fold memory reduction. Moreover, \model~ boosts accuracy by an impressive range of $1.49\%-57.18\%$ in a complex heterogeneous real-world deployment. Furthermore, we show that \model~ effectively adapts with cost-aware recomputation to dynamically shifting memory budgets. }

To the best of our knowledge, \model~ is the \textit{first} work that tackles the memory issue while taking into account the training efficiency and model performance in a unified manner. In summary, this paper makes the following key contributions:
\begin{itemize}
    \item We identify, through a comprehensive investigation of the memory footprint for model training, that the memory wall is the core constraint hampering on-device federated learning model training. In addition, by probing existing memory optimization paradigms, we define the challenges in implementing low-memory federated training frameworks on mobile devices.
    \item We design \model~, a hierarchical framework that breaks the memory wall for heterogeneous FL via model splitting. The central manager selects devices and splits layers based on memory, training, and data factors. The MEC manager handles part-training and refines splitting points, while the on-device manager optimizes local training with cost-aware checkpointing. 
    \item We conduct extensive experiments to evaluate the effectiveness of \model~ on both mobile devices and simulation testbeds with representative datasets across different models. The results demonstrate that \model~ adeptly enables memory-efficient and low-latency on-device FL training in a triply heterogeneous environment.
\end{itemize}

\section{motivation AND BACKGROUND}
\subsection{Cross-device Federated Learning.}
\begin{figure}
    \centering
    \includegraphics[width=0.8 \linewidth]{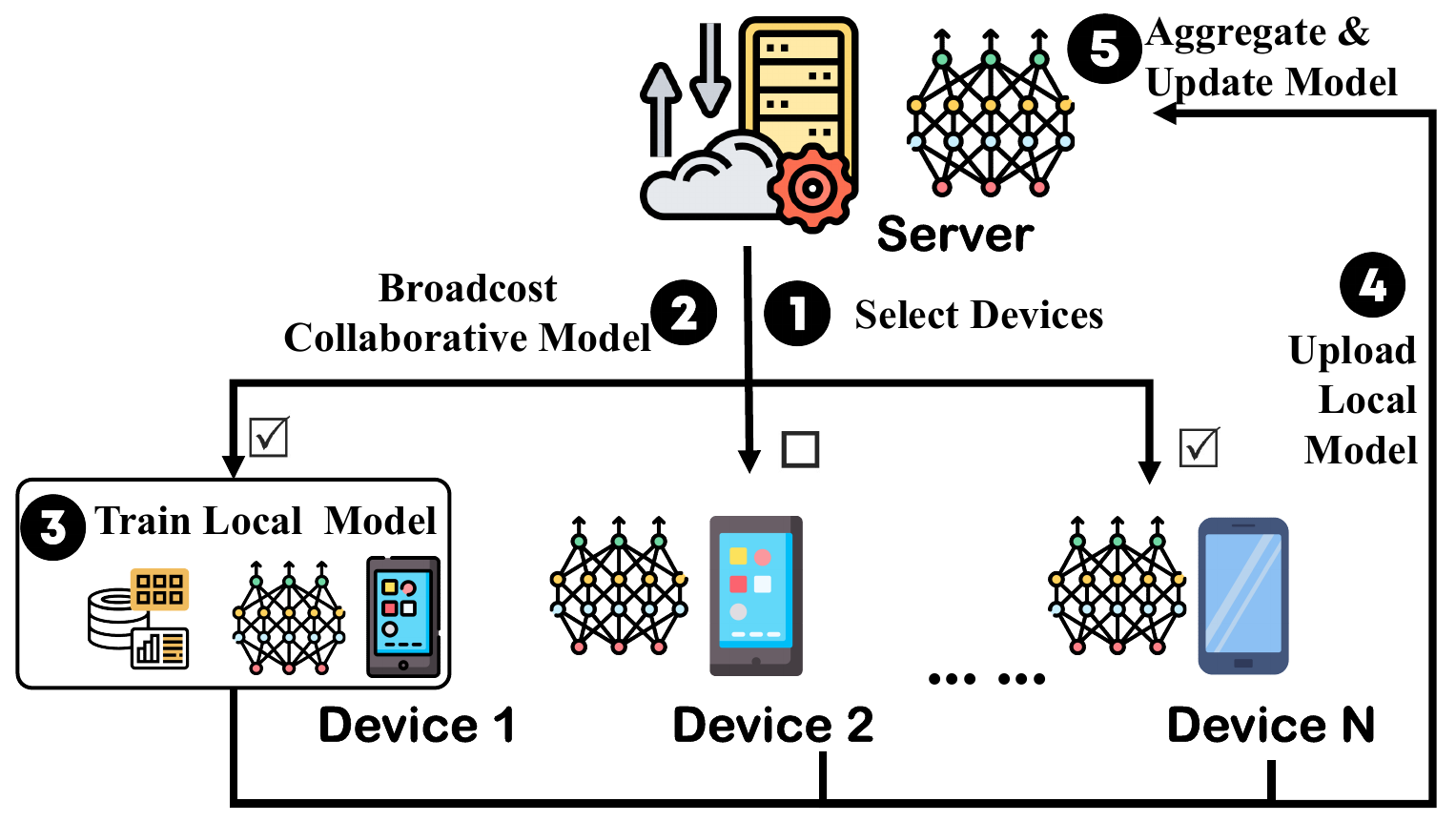}
    \caption{Workflow of Federated Learning.}
    \label{fig:FL}
\end{figure}

\begin{figure*}
    \centering
\includegraphics[width = 0.8 \linewidth]{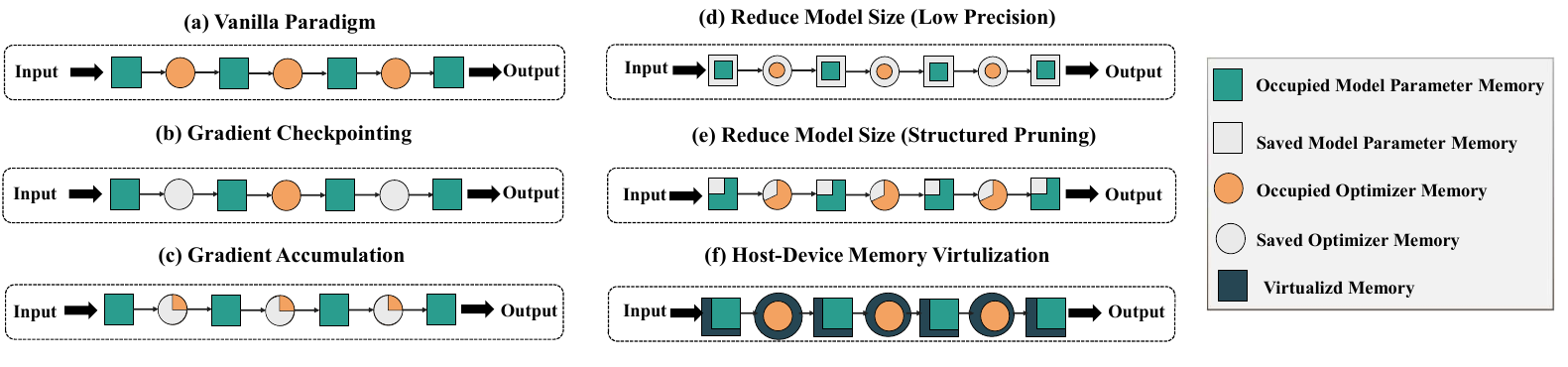}
    \caption{Diverse memory reduction strategies suitable for on-device training optimization.}
    \label{fig:memory_reduce}
\end{figure*}

Federated learning~\cite{2016federated} enables collaborative training of a shared model across multiple devices, emphasizing parallel training on local data for privacy. Figure~\ref{fig:FL} depicts the workflow of a typical federated learning framework, which unfolds as follows: \circled{1} The server randomly selects devices from the pool for each training round. \circled{2} The server sends the initialized shared model to these selected devices. \circled{3} These devices conduct local training in parallel based on their own training data. \circled{4} Subsequently, they upload their model updates to the central server. \circled{5} The server then aggregates these models to refine the shared model. This process iterates until the model converges.

\subsection{Memory Wall Hinders FL Deployment.}
\begin{figure}
    \centering
    \includegraphics[width = 0.45\linewidth]{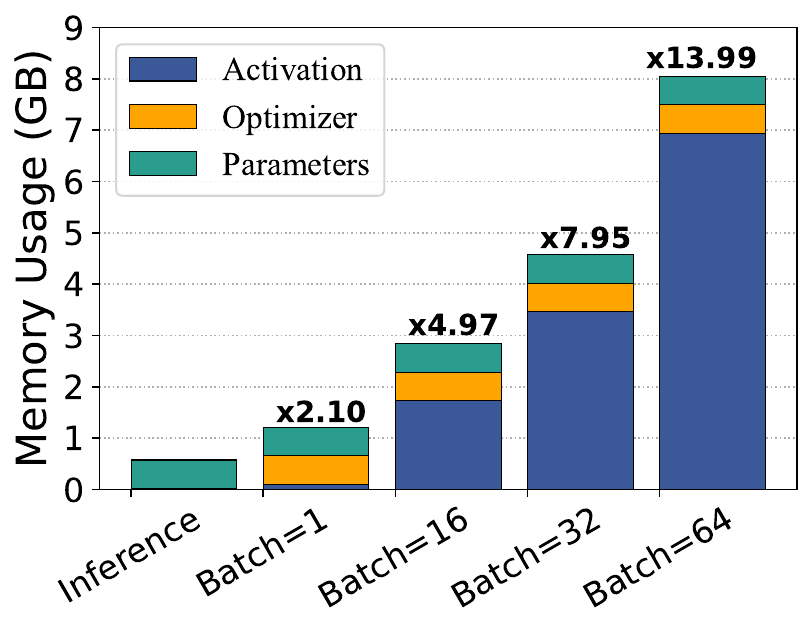}
    \includegraphics[width =0.45 \linewidth]{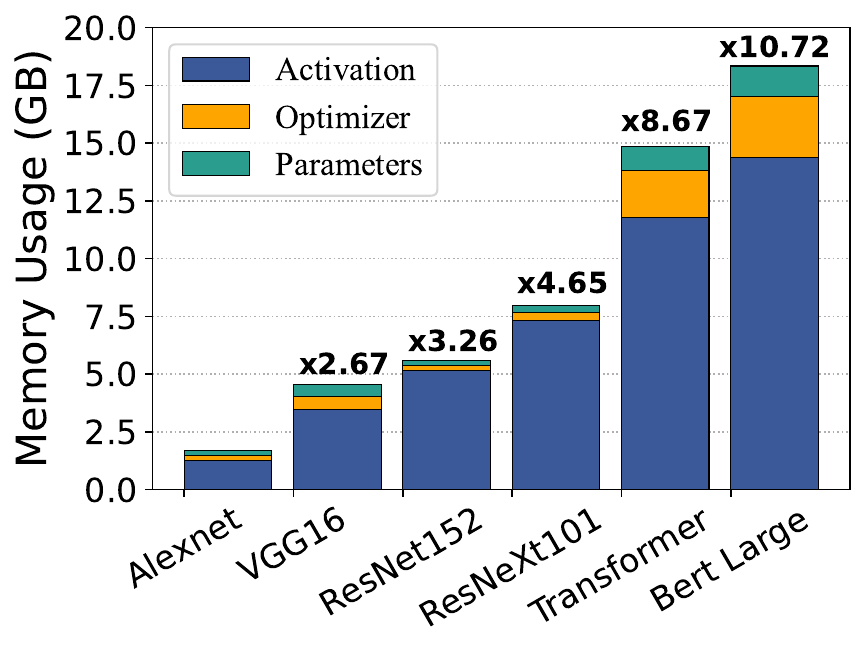} 
    \caption{Model training memory breakdown. (a) Memory comparison between inference and training for the VGG16~\cite{vgg16} model on ImageNet~\cite{imagenet} across varying batch sizes. (b) Memory analysis during device training for six commonly used models: Alexnet~\cite{alexnet} (batch size: 128), VGG16 (32), ResNet152~\cite{Resnet} (32), ResNeXt101 (32), Transformer-WMT~\cite{Transformer} (6), and Bert Large~\cite{bert} (16).}
    \label{fig:train_memory}
    \vspace{-1em}
\end{figure}

The primary hurdle in implementing real-world FL is the intensive memory requirement \cite{Add_1,Add_2,Add_3,Add_4,Add_5,Add_6,Add_7,Add_8,Add_9}. To understand, consider the essential memory components: \textit{model memory $M_{m}$} (storing layer-specific weights and biases), \textit{optimizer memory $M_{o}$} (retaining model gradients and possible momentum buffers), and \textit{activation memory $M_{a}$} (saving forward outputs and backpropagation gradients)~\cite{Low-memory}. For inference operations of the model $f(\Vec{x})$, the computational graph is linear, and intermediate states are discarded after use. Therefore, the memory required for inference is mainly influenced by the model size and the largest activation in the network, expressed as
\begin{equation}
    M_{inference} = M_{m} + \max_{i \in len|f(\Vec{x})|} M_{a_i}
\end{equation}
However, training necessitates retaining all intermediate states to derive gradients and update parameters as:
\begin{equation}
    M_{train} = M_{m} + M_{o} + \sum_{i=1}^{len|f(\Vec{x})|}M_{a_i}
\end{equation}
Figure\ref{fig:train_memory} demonstrates the notable memory consumption during both inference and training for various AI models and identifies activations as the primary memory overhead. In particular, as models grow more complex and training batch sizes increase, training memory consumption, $M_{train}$, can escalate from 5 to 100$\times$ that of inference. Compared with edge devices,
ResNeXt101 consumes about 7.96 GB, while BERT Large~\cite{bert} requires 18.34 GB. In contrast, conventional mobile devices only provide 4GB to 16GB of DRAM~\cite{RAM}. Furthermore, smartphones are limited in designating only a portion of this memory for training in order to maintain an optimal user experience. As a result, many commercial off-the-shelf mobile platforms are ill-equipped to handle the memory wall of today's deep learning models. Such a scenario highlights the necessity for a sophisticated framework that can handle FL memory limitations and ensure smooth performance on mobile platforms.

\subsection{Existing Memory Reduction Techniques}
\label{section:memory}
\begin{figure}[!ht]
    \centering
    \includegraphics[width=0.45\linewidth]{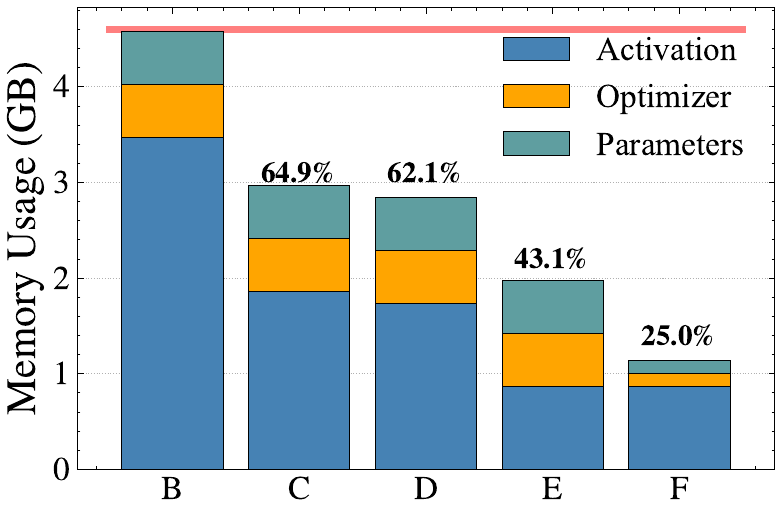}
    \includegraphics[width=0.45\linewidth]{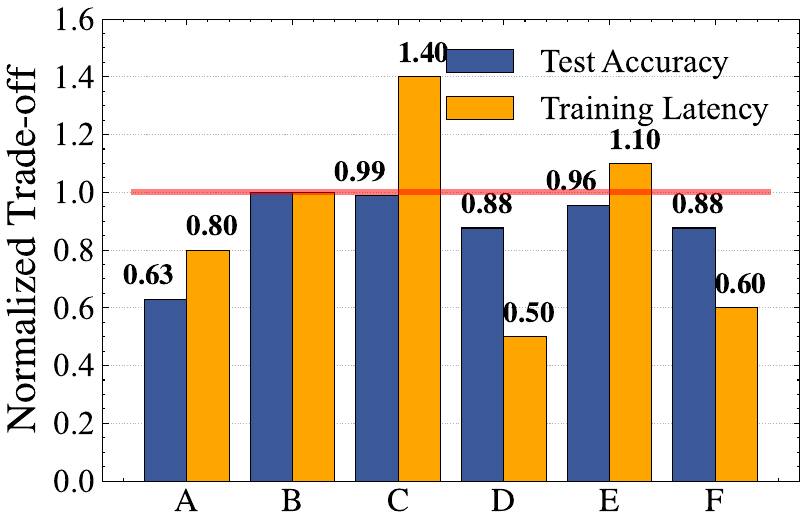}
    \caption{Profile of memory optimization techniques in federated learning, highlighting the trade-off between accuracy and training overhead. Left: memory usage. Right: model accuracy\&training time.\textbf{A}: local training with single device, \textbf{B}: vanilla FL,  \textbf{C}: gradient checkpointing, \textbf{D}: mini-batch (batch size = 16), \textbf{E}: gradient accumulation, \textbf{F}: int8.}
    \label{fig:reduce}
\end{figure}

In this section, we first explore the memory reduction techniques and then discuss why they cannot be directly utilized to support FL. Figure~\ref{fig:memory_reduce} represents the main memory optimization techniques including 1) gradient checkpointing, 2) micro-batching, 3) reducing model size, and 4) host-device memory virtualization.
To evaluate the effectiveness of these schemes in cross-device FL, we establish a pool of 100 devices and randomly select 10 devices for each training round. Each memory reduction approach is directly applied in the local training procedure. We anchor this exploration on the image classification task ImageNet with batch size 32 employing the VGG16 model. In this manner, we aim to quantitatively assess the performance, while also highlighting the advantages and trade-offs inherent in each approach, as seen in Figure~\ref{fig:reduce}.

\textbf{Gradient Checkpointing}, as shown in Figure~\ref{fig:memory_reduce} (b), selectively retains only a subset of activations in memory and recomputes non-stored activations on demand during backpropagation to obtain the gradients~\cite{checking_2, melon,Sage}. It provides a significant memory reduction in training compared to the vanilla paradigm shown in Figure~\ref{fig:memory_reduce} (a). In particular, as shown in Figure~\ref{fig:reduce}, employing the checkpointing technique in federated learning, as per~\cite{checking_1},  only consumes 64.9\% of memory without compromising model accuracy. However, this gain is offset by increased computational overhead during local training, which increases training time by approximately 1.4$\times$. It severely aggravates the straggler problem caused by hardware heterogeneity and slows down the overall training progress.

\textbf{Micro-batching} is a technique that minimizes the batch size, effectively reducing the activations in each training iteration and thereby alleviating memory pressure. While using smaller batches seems intuitive, it can lead to inaccurate global gradient estimations during the descent phase, risking oscillatory losses or model non-convergence. To more accurately capture the global gradient, the approach of gradient accumulation has been introduced~\cite{accumulate}. As depicted in Figure~\ref{fig:memory_reduce} (c), gradients from consecutive micro-batches are gathered in a specialized buffer. Once enough data is collected, updates to both the model parameters and momentum buffer are made, after which the gradient buffer is reset. We implement a micro-batch approach (with a batch size of 4) in FL. Figure~\ref{fig:reduce} shows that it can yield a memory reduction of up to 43.1\%. It achieves memory reduction at the expense of decreased parallelism. While this doesn't alter the total computational requirement or hinder model accuracy, it leads to a 1.1$\times$ increase in training time, which could affect overall system efficiency.

\textbf{Reducing Model Size} is a strategic approach to alleviate memory burden by compacting the model's representation or structure. This incorporates techniques like model pruning~\cite{pruning}, low-rank factorization~\cite{lowrank}, quantization~\cite{mixed,int_2,int_3}, knowledge distillation~\cite{KD}, and compact model design. Specifically, \textit{Quantization}, illustrated in Figure~\ref{fig:memory_reduce} (d), leverages lower precision to represent model parameters and activations, conserving memory. On the other hand, \textit{Pruning}, depicted in Figure~\ref{fig:memory_reduce} (e), zeroes in on eliminating less impactful network connections, further downsizing the model. In FL applications, as depicted in Figure~\ref{fig:reduce}, employing techniques like quantization (Int8) demonstrates notable memory efficiency that uses merely 25\% of typical memory. This not only eases memory demands but also shortens training time due to computational efficiencies. However, quantization decreases model accuracy by 12\%, which seriously compromises the model performance.

\textbf{Host-Device Memory Virtualization}, depicted in Figure~\ref{fig:memory_reduce}  (f), extends the logical memory of the GPU in server configurations by utilizing host-side memory. It offers a promising solution for large-scale server-side cross-silo model training on servers~\cite{vitrual_1}. Mobile SoCs, however, represent a different architecture: these compact integrated circuits combine computational elements such as CPUs, GPUs, and memory controllers on a single chip and operate under a unified memory scheme where CPUs and GPUs share the same physical memory. While this shared memory architecture streamlines data transfers and improves operational efficiency, it also introduces potential contention between the CPU and GPU during concurrent data access, especially in intensive model training scenarios. Thus, the intrinsic shared memory design of mobile SoCs, compounded by other inherent resource constraints, makes the straightforward adoption of such virtualization techniques in mobile contexts a complex endeavor.

In summary, the existing memory optimization techniques reduce the memory requirement during the local training process at the expense of either training progress or model accuracy. They cannot be directly utilized to support FL in practical settings

\subsection{Shined by Model Splitting}
\label{challenge}

\begin{figure}[!ht]
    \centering
    \includegraphics[width =0.8\linewidth]{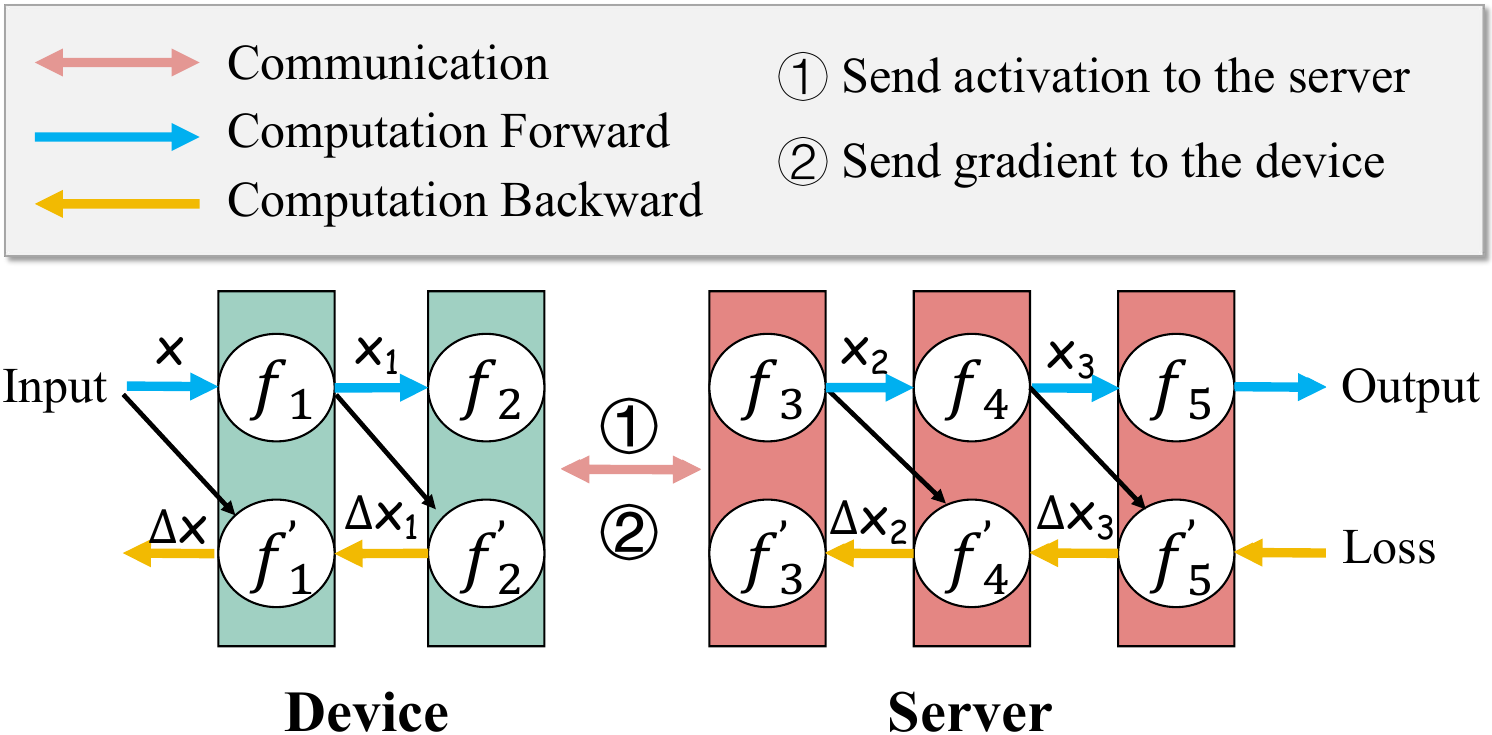}
    \caption{\tian{During each training round, the client computes the forward propagation of its client-side model in parallel and sends activations to the server. The server then computes the forward and backward propagation of the server-side model and sends the gradients (corresponding to the client-side activations) back to the client.}}
    \label{fig:fedsplit}
\end{figure}

Model splitting~\cite{split_2,split_3,split_6} shines the light on breaking the memory wall of local training. Figure~\ref{fig:fedsplit} represents the training process with model splitting. It splits a deep learning model into two portions, the device-side network and server-side network. In the forward path, the activations of the split layer are sent to the server, while the gradients are received during backpropagation. The memory requirement on the device side can be effectively reduced through model splitting. However, breaking the memory wall of heterogeneous FL with model splitting still faces the following critical challenges. 

\subsubsection{Triple Heterogeneity} An FL system with model splitting faces heterogeneity from the following three perspectives. First, the clients have totally different computing capabilities and memory constraints due to hardware heterogeneity. Second, the training data across various clients are non-IID. In addition, for the shared model, each layer consumes highly different computing and memory resources. Thus, the selection of the participating clients and the cut layer can highly impact memory consumption, training efficiency, and model accuracy at the same time. Thus, the first challenge is how to jointly handle the triple heterogeneity in a unified manner.

\subsubsection{Communication Bottleneck}
\tian{When conducting local training with model splitting, the edge device needs to send the activations and receive the gradients from the central server during each training iteration.} It causes high communication overhead and decelerates the overall training progress. Thus, the second challenge is how to overcome the communication bottleneck caused by model splitting.

\subsubsection{Dynamic Memory Budget} 
 Due to the resource contention caused by the concurrently running apps, the memory budget for the training process can be highly dynamic. Thus, how to guarantee the local training procedure proceeds successfully with a highly dynamic memory budget is the third critical challenge.

\section{High-level Ideas of \model~}
In this section, we first introduce the system and statistical models and then discuss how \model~ jointly considers memory reduction, model accuracy, and training efficiency in a unified manner. 

\subsection{System Utility}
Neural network computation can be expressed as a directed graph $G = (V, E)$, where $V = \{v_1, u_2, ..., v_N\}$ is the set of nodes, each representing a layer in the neural network, and $E \subseteq V\times V$ is the set of edges. Each directed edge $(u, v)$ represents the dependency that the operation of $v$ cannot be executed until that of $u$ is finished. We use $Pr(v) = \{u: (u, v) \in E\}$ to represent the set of immediate predecessor nodes of $v$.
To model the runtime of a layer $u$, we decompose the execution time into three terms as a simple summation:
\begin{equation}
T^{cp}(u) = R(Pr(u)) + C(f_u,d) + W(f_u,d)
\end{equation}

where $R(Pr(u))$ is the IO time to fetch the input produced by its predecessor layers.  $W(f_u,d)$ is the time to write the outputs to the local memory. They are calculated as the amount of memory involved in the computation divided by the IO bandwidth of the device. For model split learning,  the inputs must be fetched from other devices, therefore this IO bandwidth refers to the communication bandwidth between the server and the local device. $C(f_u,d)$: the computation time to perform the neural network computation of $f_u$ on the specified device $d$, which is calculated as the FLOP (floating-point operation) counts of the operation divided by the computation speed (FLOPS, or floating-point operation per second) of the device: $C(f_u,d)=\frac{FLOPs(f_u)}{speed(d)}$.
Based on the per-layer computation model, we can model the training process latency of an entire network of device $i$ as:
\begin{equation}
\begin{split}
T = \sum_{u=1}^{j}T^{D}(u) + \frac{O(j)}{r} + \sum_{u=j+1}^{V}T^{E}(u)
\end{split}
\end{equation}

where $O(\cdot)$ refers to the output activation size of the split layer. $T^{D}(\cdot)$ and $T^{E}(\cdot)$ denote processing sub-network on the mobile device and server, respectively. V refers to the number of model layers.

However, for FL, reducing the processing time of an individual device does not ensure the efficiency of the whole system. High-end mobile devices still have to wait for other devices (i.e., stragglers, low-end mobile devices with low computing power and low communication bandwidth) to complete the training process. The training latency for each round can be denoted as:
\begin{equation}
    T_{system} = \max_{i \in N} \{T(i,j)\}
    \label{eq:sys}
\end{equation}

\subsection{Statistical Utility}
In order to model the data distribution across different clients, we adopt the Kullback–Leibler Divergence (KLD)~\cite{KLD_1,KLD_2,KLD_3} to quantify the balanced degree of data distribution of device data distribution. The smaller KLD value means the data distribution is closer to the uniform distribution. Note that, in this paper, the data distribution
of each node can be unknown ahead, which can be effectively inferred from the uploaded model updates without sacrificing data privacy based on the previous work~\cite{data_dis_1,data_dis_3,HARMONY}.
\begin{equation}
   \label{eq:dis}
   \operatorname{Dis}(i) = D_{K L}\left(P_d \| P_{exp}\right)= \sum_{i=1}^{N} P_d\left(p_j\right) * \log \frac{P_d\left(p_i\right)}{P_{\exp}\left(p_i\right)}
\end{equation}
where $P_d$ is the local device's data distribution, while $P_{exp}$ is the expected uniform distribution.

As the training process proceeds, the contribution of the same sample to model training evolves. Excluding non-important samples can further reduce unnecessary training overhead. We argue that a device that has a higher training loss on its local dataset should have a higher statistical utility~\cite{fedbalancer, SampleSelection1, SampleSelection2}.
\begin{equation}
\label{eq:stat}
\operatorname{Stat}(i) =\left|D^i\right| \sqrt{\frac{1}{\left|D^i\right|} \sum_{s \in D^i} \mathcal{L}(s)^2}
\end{equation}
where $D^{i}$ means the locally stored training samples on device $i$ and, and $\mathcal{L}(s)$ represents the training loss of the local model got from sample record $s$.
Intuitively, this statistical utility ranges from 0 to $\infty$, depending on the loss function. A high value means higher data gains. Composing the above importance and distribution model, we design the data utility as $U_{data}= f(\operatorname{Dis}(i), \operatorname{Stat}(i))$.

\begin{figure}[!t]
    \centering
    \includegraphics[width = 0.8\linewidth]{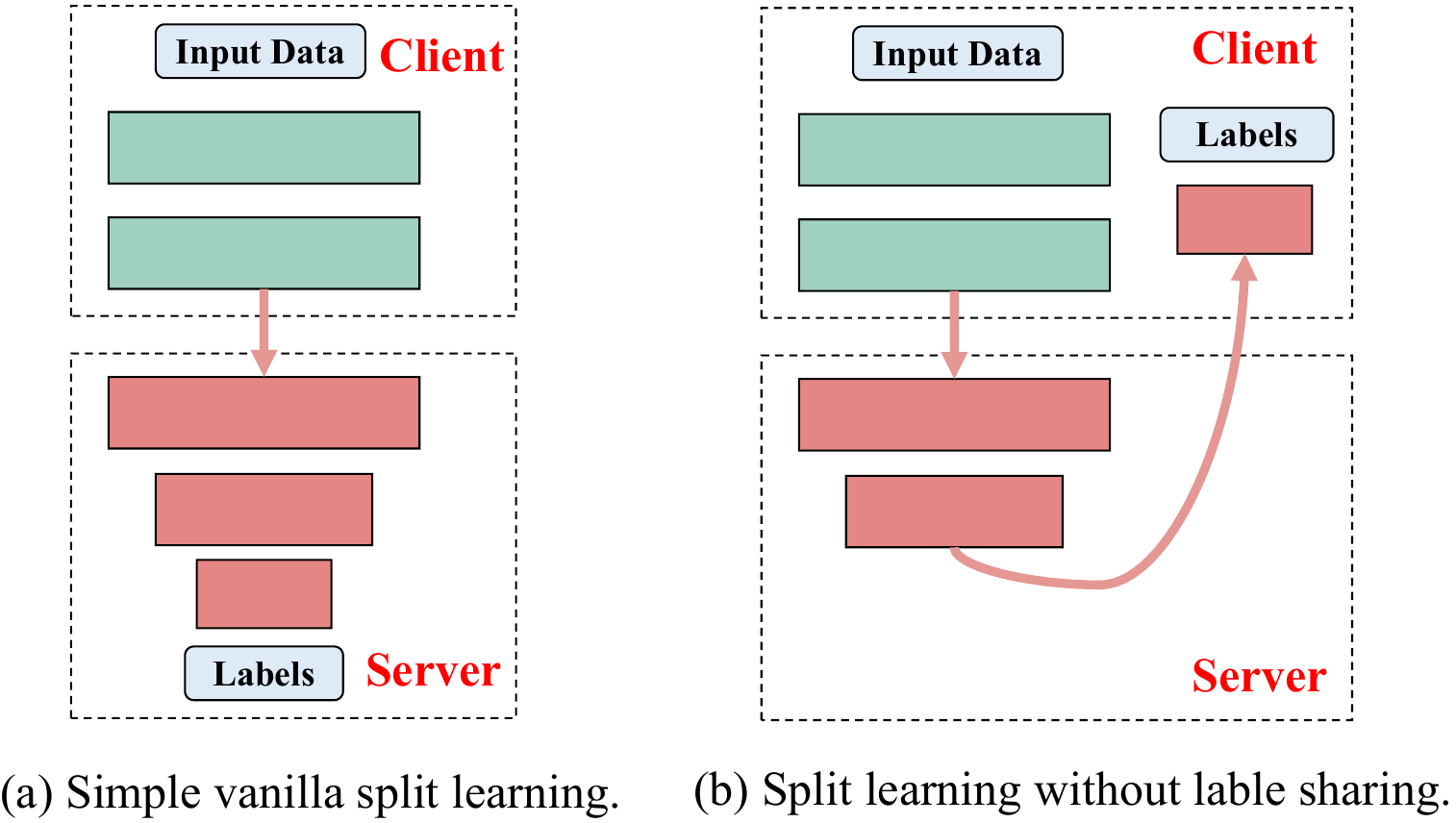}
    \caption{\tianNew{Two architectures for split learning. (a) Vanilla structure: Existing split federated learning framework~\cite{splitfed, split_2,split_3,split_4,split_5,split_6,fedadapt,split_d_2,split_d_3}, the training labels hosted in the server may cause privacy leakage. (b) U-split structure that keeps labels with sensitive information on the local device. To protect privacy well, the second one is adopted in this work.}}
    \label{fig:split_framework}
\end{figure}

\subsection{\tianNew{How \model~ respects privacy?}}
\tianNew{
Exploring the privacy-sensitive aspects of device information (e.g., raw data, labels) may discourage participants from contributing to FL \cite{2016federated}. Therefore, practical FL systems must improve efficiency while operating with the limited information available in real-world applications, and their deployment needs to be lightweight for devices. To protect end-user privacy, we have designed the following modifications:}

\tianNew{
\begin{enumerate}
    \item \textbf{U-split structure}. As depicted in Figure \ref{fig:split_framework}, unlike conventional federated split learning architectures employed by most existing methods~\cite{split_2,split_3,split_6}, which store training labels on the server and thereby pose significant privacy risks by exposing sensitive user information, our U-split framework retains labels containing sensitive information on the local device, enhancing privacy protection.
    \item \textbf{Upload Information}. Training loss, which quantifies the model's prediction confidence without disclosing raw data, is frequently collected in real-world FL deployments \cite{  fedranking}. We aggregate the training loss computed locally by the client over all samples, ensuring that the loss distribution of individual samples remains masked. To further enhance privacy, clients can add noise~\cite{geyer2017differentially} to their loss values prior to uploading, providing theoretical privacy guarantees for SmartSplit. Additionally, sending the training loss to the central server avoids introducing extra processing latency and communication costs, since the probing states (scalars) are tiny compared to the DNN model.
\end{enumerate} 
}

\subsection{Problem Formulation}
To unify the system model $T_{system}$, and the data model $U_{data}$ for device and splitting point selection, we formulate the partitioning problem from the global perspective as a heterogeneity-aware utility function as follows:
\begin{equation}
F(S) = \underset{(i\in N,j\in V)}{\operatorname{argmin}} \sum_{\forall \alpha_{i}=1} f\left( T_{system}, U_{data}\right) 
\end{equation}
s.t:
\begin{subequations}
\begin{align}
	1 &\le i\le N \nonumber\\
    1 &\le j \le V \nonumber\\
	\alpha_{i} &\in \left \{ 0,1 \right \} \nonumber\\
	\sum_{i\in N}^{}\alpha_{i}*D^{i} &\ge D_{threshold} \nonumber\\
    \sum_{0 \le u \le j} M_{i}(u) &\le M_{i, budget} \nonumber
\end{align}
\end{subequations}

The solution of the problem yields a selected indicator matrix $S \in R_{N \times V}$, where $S_{i,j}=(\alpha_{i},v_{i,j})$, means selecting efficient devices and corresponding optimal splitting scheme to build the whole system well-orchestrated. $\alpha$ ensures that the state indicator should only be 1 for being selected or 0 for not being selected. $D_{threshold}$ is a developer-specified size of the training set to guarantee model accuracy. $M_{budget}$ is the available memory budget for the device.

\section{\model~: System Design}
In this section, we begin with an overview of the system, followed by a detailed description of each component.

\subsection{System Overview}
\begin{figure}
    \centering
    \includegraphics[width= 0.8\linewidth]{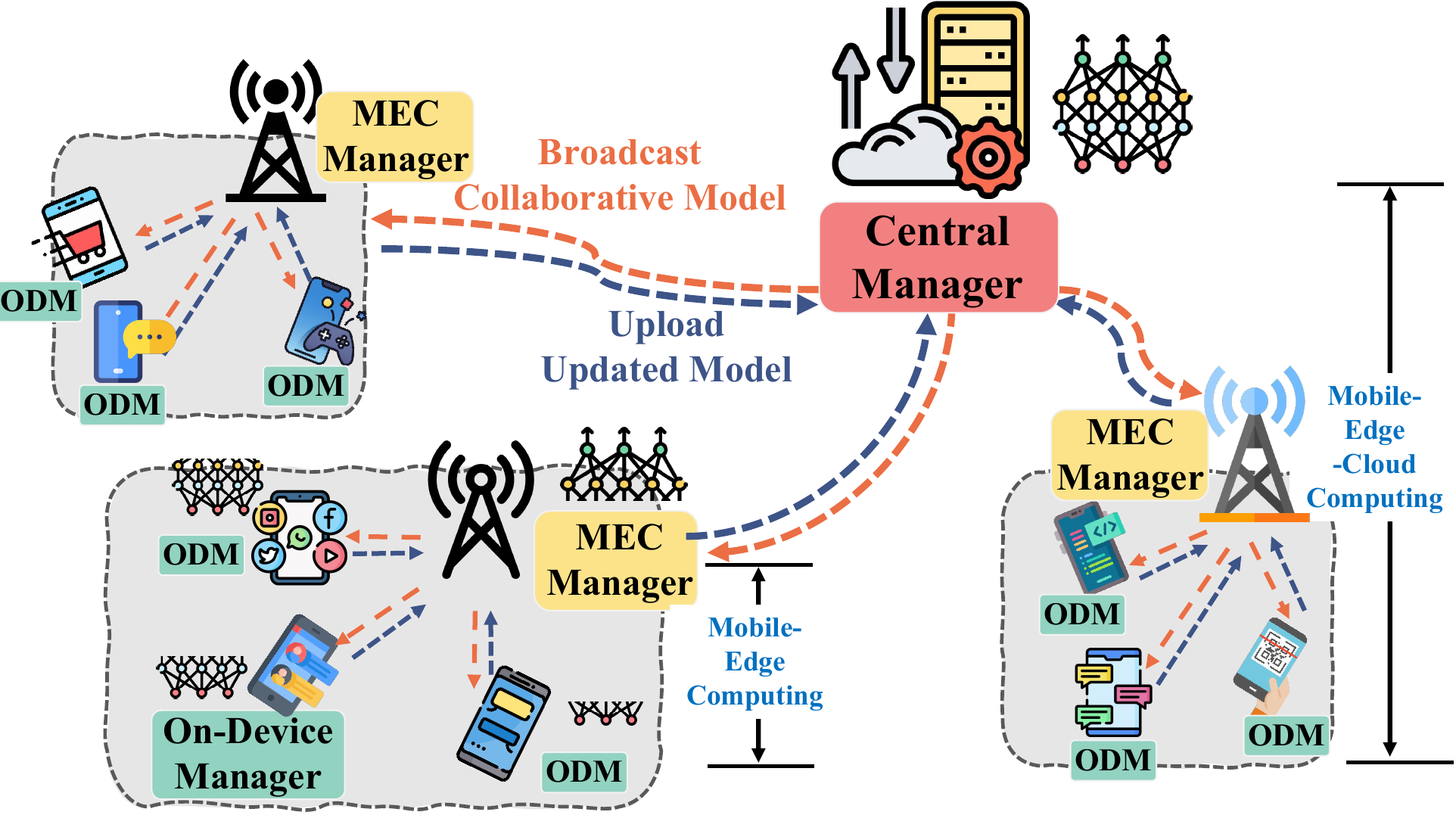}
    \caption{\tian{System architecture of the \model~. During training, the central manager selects devices, splits models, and directs the MEC manager to deploy models on devices and servers. This minimizes communication overhead and enhances efficiency through dynamic re-selection and monitoring, ensuring effective training until convergence.}}
    \label{fig:sys_arch}
\end{figure}

\tian{\model~ employs a three-tiered, Device-MEC-Server, hierarchical management system.} The system architecture is illustrated in Figure~\ref{fig:sys_arch}, that \textit{Central Manager}, \textit{MEC Manager}, and \textit{On-Device Manager (ODM)} are embedded in each tier. In each training round, the central manager first conducts device selection and determines the splitting point of the local models for the selected devices according to the runtime training capability and data distribution across different devices. The selection results are then sent as input to the MEC manager. After that, the MEC manager partitions the local models of the selected devices into two parts. The server part is deployed on the MEC server while the device part is deployed on the corresponding selected device. Local training with model splitting is then conducted between the mobile device and the edge server. For the reason that the edge server is much closer to the devices, the communication overhead during the training process is effectively mitigated. Meanwhile, the MEC manager re-selects the participating devices based on the time-evolving data importance to further improve the training efficiency. At the same time, the On-Device Manager probes the device's runtime status training capacity, memory budget, and loss. Concurrently, the on-device Manager adopts cost-aware checkpointing to balance memory and training efficiency. The updated local models are then sent to the central server for model aggregation. After that, the whole system enters the next training round till the model converges. 

\subsection{Central Manager}
The central manager, as the global coordinator of the system, collects the runtime information including conducting client selection and splitting layer determination for each local model in each training round. Then, the central server broadcasts the model and selection matrix to the MEC. 

\tian{We develop the Bayesian Optimization (BO)~\cite{BO_1,BO_2,BO_3} model to effectively manage model splitting across a network of devices, characterized by varying computational capabilities and statistical utility. BO operates as a black-box optimization strategy that excels in dealing with complex, unknown utility functions. It systematically explores and exploits the search space $\chi$, which represents the combination of potential split-points in the computational model (represented by $V$) and potential participant device number (represented by $N$). This exploration is guided by evaluations of the utility function across various configurations of model splits, aiming to achieve an optimal distribution of computational loads.}

\tian{
The optimization process involves the construction of a stochastic model that approximates the performance or utility of different model splits based on observed data points. As new data is gathered through trials at different configurations (i.e., different allocations of model segments to devices), the stochastic model, typically a Gaussian Process, is continuously updated to better reflect the underlying utility function. The objective is to minimize the overall system latency and computational load. Specifically, based on the training time $T_{system}$, according to Eq.~\eqref{eq:sys}, which measures the training capacity of local devices, and the KLD value $Dis$ of estimated raw data distribution, according to Eq.~\eqref{eq:dis}, the optimal devices and corresponding split-layer selection within the memory budget $M_{budget}$ can be represented as follows:
 \begin{equation}
     F(S) = \min_{\alpha_{i}} \sum_{i=1}^{N}[\alpha*(Dis(i)+\lambda*T_{system})]
 \end{equation}
where $\lambda$ is a regularization parameter that balances distribution cost against the system latency $T_{system}$. The acquisition function within BO is used to determine the most promising next sample point $x^*$ in the search space $\chi$, which corresponds to a potential new model split configuration. This is formally described by:
\begin{equation}
    x^* =\mathop{argmax} \limits_{x\in \chi}  f(x)
    \label{Eq.1}
\end{equation}
where $f(x)$ is the objective function modeling the utility of each configuration, aiming to identify the configuration that yields the highest utility with the fewest possible evaluations. $\chi$ is effectively the two-dimensional space $V \times N$, while $x^{*}$ represents the indicator matrix $S$ that specifies the optimal device and split-point selection, thus enabling efficient computation distribution and reduced latency across the network. It not only optimizes the allocation of computational tasks but also adapts to dynamic changes in device availability and network conditions, ensuring robust performance in distributed computing environments.}

\subsection{MEC Manager}
MEC manager, serving as a top-down orchestrator, is instrumental in diminishing communication overhead during model splitting. \tian{The MEC Manager accelerates training by collaborating with local devices to manage load and avoid delays from slow performers. This setup not only resolves server overload and conflict issues but also ensures that delays do not affect the overall training schedule. In the proposed three-tier Device-MEC-Server architecture, depicted in Figure \ref{fig:fedsplit} and Figure \ref{fig:sys_arch}, the MEC Manager reduces communication overhead significantly by optimizing data flow within a high-bandwidth local area network (LAN). Additionally, as training evolves, the MEC Manager updates loss metrics and the Importance Scheduler adjusts data prioritization, enhancing training efficiency and model effectiveness.}

\begin{figure}
    \centering
    \includegraphics[width =0.8\linewidth]{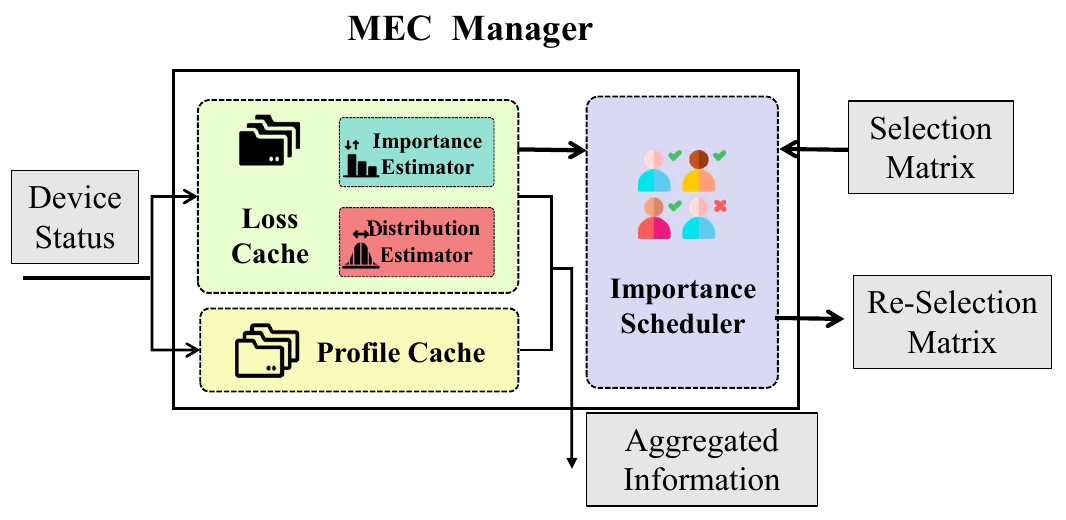}
    \caption{Architecture and workflow of MEC manager.}
    \label{fig:MEC}
\end{figure}

As illustrated in Figure~\ref{fig:MEC}, it utilizes the system profile and probing loss of mobile devices within the jurisdiction as inputs, and stores these in the \textit{Profile Cache} and \textit{Loss Cache} respectively. After that, caches guide the \textit{Importance Estimator} and \textit{Distribution Estimator} as efficiency assessments. Finally, the MEC manager forwards aggregated information to the central manager. Simultaneously, it supervises the ongoing training process and updates device loss metrics, while the \textit{Importance Scheduler} re-calibrates data importance for re-selection for future iterations. Upon completing local iterations, the MEC manager uploads the updated model to the central server.

\textbf{Loss \& Profile Cache.}
The loss cache stores the probing loss to estimate the data distribution and importance. The profile cache stores the system profile of governed devices. As described in Alg.~\ref{alg:estumator}, \textit{Distribution Estimator} utilizes the probing weight with help of tiny public auxiliary dataset $D_{aux}$ ($<<\sum_{i \in N}D_{i}$) to obtain the approximate data distribution~\cite{data_dis_3,data_dis_4} (Line~\ref{1_2},\ref{1_6}-\ref{1_10}), while estimator KLD value of this distribution (Line~\ref{1_3}), according to \eqref{eq:dis}. Meanwhile, the \textit{Importance Estimator} employs the \eqref{eq:stat} to produce the training sample importance (Line~\ref{1_4}). 

\SetKwFunction{proca}{Obtain\_{dis}}
\SetKwProg{proc}{Function}{}{}

\begin{algorithm}
\caption{Distribution \& Importance Estimator}
\label{alg:estumator}

\KwData{Auxiliary dataset $D_{aux}$, probing model $W_{i}$ and probing loss $\mathcal{L}_{i}$ of device $i$.}
\KwResult{Data distribution KLD value $Dis(i)$, and data importance value $Stat(i)$.}

\For{i $\in$ N}{$P_{i}$ $\leftarrow$ Obtain\_dis ($D_{aux}$,$W_{i}$)\; \label{1_2}
$Dis(i)$ $\leftarrow$ Dis\_model($P_{i}$); \textcolor{gray}{$\rhd$ Based on \eqref{eq:dis}}\\ \label{1_3}
$Stat(i)$ $\leftarrow$ Imp\_model($\mathcal{L}_{i}$); \label{1_4} \textcolor{gray}{ $\rhd$ Based on \eqref{eq:stat}}}
\proc{\proca{$D_{aux}$, $W_{i}^{r}$}}
{\label{1_6}
$\nabla \mathcal{L}(W_{i})$ $\leftarrow$ ($ W_{i}$, $D_{aux}$) \;
\For{each class in parallel}{$P_{i}$ $\leftarrow$ $P_{i} \cup Class\_ratio(p_{j})$\;}
\label{1_10}
}
\end{algorithm}

\textbf{Importance Scheduler.}
80/20 Rule~\cite{fedbalancer} states partial resources are occupied by a few users. Therefore, at each training epoch, the importance scheduler re-selects participating devices according to the estimated sample importance. Then gradually removes $\sigma$\% already-learned samples to speed up training and reduce computational resources, where $\sigma$\footnote{Note that we use 80\% for $\sigma$ based on previous works~\cite{fedbalancer}.} is a developer-specified ratio. Alg.~\ref{alg:fine} summarizes the scheduling process, if the random value evaluated by the scheduler is lower than specified $\epsilon$\footnote{We set $\epsilon$ as 0.1 based on sensitivity analysis.} 
\model~ will randomly select participants. Otherwise, it will sort the device data importance utility by $Stat$ and select the top $K$ devices. After the scheduling is finished, MEC accesses the training process and updates the values in the loss cache after each training iteration.
\begin{algorithm}
\caption{Importance Scheduling }
\label{alg:fine}
\KwData{Data importance value $Stat(i)$, exploration probability $\epsilon$, split policy indicator $S$.}
\KwResult{Fine-selected devices $N_{fine}$.}
$U_{stst} \leftarrow \emptyset$; \textcolor{gray}{$\rhd$ Logging devices importance}\\
\While{During the local training iteration}{
\For{i $\in$ S}{$U_{stat}.append(Stat(i)$)}
\lIf{Rand() $<$ $\epsilon$}{\\ \quad Choose $K$ devices randomly}
\lElse{\\ \quad Sort devices by $U_{stat}$\;
        \quad Choose at most top $K$ devices}
\textbf{end }\\
MEC and On-Device training\;
Update the Loss Cache\;
}
\end{algorithm}

\subsection{On-Device Manager}
\begin{figure}
    \centering
    \includegraphics[width= 0.8\linewidth]{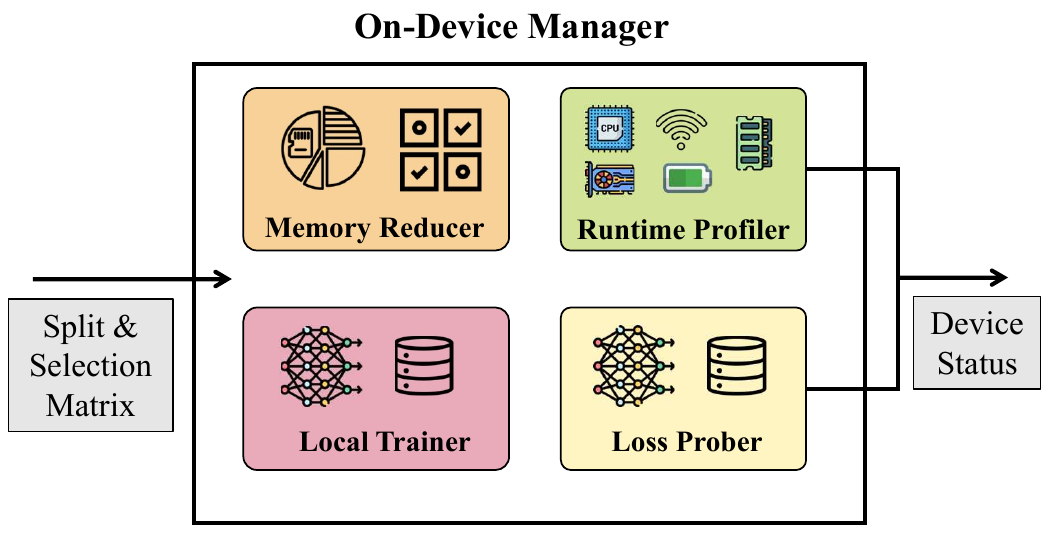}
    \caption{\tian{Architecture and workflow of on-device manager (ODM), embedded in each mobile device, strategically coordinates training by monitoring local data and managing memory to optimize usage without sacrificing efficiency. During initialization, the ODM uses the Runtime Profiler and Loss Prober to assess system capacity and training loss, while the Memory Reducer actively reduces memory overhead within a dynamic budget. This approach helps balance memory efficiency with computational needs, boosting system performance.}}
    \label{fig:mobile}
\end{figure}

The on-device manager is integrated into each mobile device with twin objectives: 1) monitors local data and system profiles for holistic training coordination, and 2) discerns the dynamic runtime memory budget, enabling memory optimization during model training.
Notably, deducing an approximate data distribution without breaching privacy and trimming the memory usage without compromising model efficiency remain formidable challenges. \tian{As depicted in Figure~\ref{fig:mobile}, during the initialization phase, the ODM employs both the \textit{Runtime Profiler} and \textit{Loss Prober} to measure system capacity and training loss respectively. Subsequently, the \textit{Memory Reducer} minimizes training memory overhead within the dynamic memory budget.}

\textbf{Memory Reducer.}
Resource contention induces dynamic memory budget on mobile devices. To address this, we release the tensors in cheap-to-compute for reconstructions, i.e., checkpointing. Generally, there are two methods for memory recomputation segment: memory-centric and speed-centric~\cite{Superneurons}.

The speed-centric approach recomputes a segment $seg$ only once, with subsequent layers reusing these tensors. Thus, it only incurs $O(N)$ extra computations, and the memory cost $M_{cost}$ is expressed as:
\begin{equation}
M_{cost} = \sum_{i=1}^{seg} l_{i}^{f} + l_{seg}^{b}
\label{eq: speed-centric}
\end{equation}
where $l_{i}^{f}$ and $l_{i}^{b}$ mean forward and back-propagation memory usage of layer $i$ respectively. For instance, for the backward step on $l_{4}^{b}$, dependencies $l_{1}^{f}$ $\rightarrow$ $l_{3}^{f}$ are recomputed. It retains $l_{1}^{f}$, $l_{2}^{f}$ for reuse in the backward computations of $l_{3}^{b}$ and $l_{2}^{b}$. However, if the peak memory layer $l_{peak}$ within the segment, $M_{cost}$ will exceed $l_{peak}$. The memory-centric method recomputes forward dependencies for each backward layer, releasing intermediate results to maximize memory savings $M_{cost}=l_{i}^{b}$, never exceeding $l_{peak}$, but at the cost of $O(N^{2})$ additional computations. For instance, while computing $l_{4}^{b}$, it recomputes $l_{1}^{f}$ $\rightarrow$ $l_{3}^{f}$; and for $l_{3}^{b}$, it recomputes $l_{1}^{f}$ $\rightarrow$ $l_{2}^{f}$. 

In this work, we employ cost-aware recomputation, merging the benefits of both methods that if $\sum_{i=1}^{seg} l_{i}^{f} + l_{seg}^{b} \le l_{peak}$, the speed-centric method is applied, otherwise, the memory-centric one is used.
Cost-aware recomputation ensures $M_{cost} \le l_{peak}$ to be consistent with the memory-centric strategy making the network-wide  $M_{budget} = l_{peak}$, while the extra computations are comparable to the speed-centric strategy. As Alg.~\ref{alg:memery} shows, the memory reducer iterates over all the model layers to find peak memory usage one $l_{peak}$ (Line~\ref{Line: peak}). This intelligent cost-aware recomputation scheme first verifies that there is enough memory available for a speed-centric computation (Line~\ref{Line: peak_1}). If the memory budget is sufficient, the speed-centric approach is applied (Line~\ref{Line: peak_2}). Otherwise, a memory-centric strategy prevails (Line~\ref{Line: peak_3}).

\begin{algorithm}
\caption{Memory Reducer}
\label{alg:memery}
\KwData{Forward/back-propagating memory at layer $i$: $l_{i}^{f}$ and $l_{i}^{b}$, peak memory layer $l_{peak}$, recomputation segment set $S$, memory cost $M_{cost}$.}
\KwResult{Recomputation method: $R$.}

$l_{peak} \leftarrow max(l_{i})$;  \label{Line: peak}\\
\For{seg $\in$ $S$ \label{Line: peak_1}}{
$M_{cost}^{b} = \sum_{i=1}^{seg} l_{i}^{f} + l_{seg}^{b}$\\
\lIf{$M_{cost}^{b} \le M_{peak}$}{\\ \quad $R \leftarrow \text{speed-centric}$ \label{Line: peak_2}}
\lElse{\\ \quad $R \leftarrow \text{memory-centric}$ \label{Line: peak_3}}}
\end{algorithm}

\textbf{Runtime Profiler \& Loss Prober.}
From a system perspective, the profiler monitors the ongoing training state. Specifically, it measures memory budget and measures processing speed (FLOPS), among other key metrics.
From a data perspective, we start probing training in the first local training epoch to obtain loss that adeptly encapsulates the raw data characteristics (distribution and importance). Then, they output the device status to the superior manager.

\section{Evaluation}
In this section, we present our empirical evaluation of \model~.
\subsection{Experiment Setup}

 \textbf{Infrastructure.}
As a proof-of-concept case study, we built the following testbed to faithfully emulate the triple heterogeneity real-world deployment configurations.
First, we built a simulator with the server-edge-device structure using PyTorch~\cite{pytorch}, which deploys different processes to emulate the central server, MECs, and corresponding mobile devices. 
Second, using Dirichlet distribution~\cite{dd_1,dd_2,dd_3} to construct different distributions are assigned as heterogeneous statistical setup.
\tian{Third, as shown in Table \ref{tab:device_hardware}, we employed five types of mobile devices with different SoCs and memory configurations to build an FL system, including Redmi2, Honor8, Redmi Note10, Google Pixel6, and Oneplus9, as the heterogeneous resource setup.} \tian{We adopt the Dell 5820 tower workstation with NVIDIA RTX3090 GPU paired 24 GB GDDR6X memory as the MEC server. We utilized Deep Learning4java (DL4J)~\cite{DL4J} as a background service to implement the on-device learning, while using a Monsoon Power Monitor~\cite{Monsoon} to measure the power consumption of the training process.} We then set up a FL system consisting of 100 devices, 5 MECs, and a server.

\begin{table}[!t]
	\caption{\color{black}{Hardware specifications of the user-end device.}} 
	\centering 
	\resizebox{0.95\linewidth}{!}{ 
	\begin{tabular}{c|c c c} 
		\toprule[1.5pt]
		\rowcolor{gray!20} \color{black}{\textbf{Hardware}} & \color{black}{\textbf{SoC}} & \color{black}{\textbf{CPU (GHz)}} & \color{black}{\textbf{RAM (GB)}}\\ 
		\midrule[0.75pt]
		\color{black}{Redmi2} & \color{black}{Snapdragon 410} & \color{black}{4 $\times$ 1.20}  & \color{black}{2}\\ 
		\hline
		\color{black}{Honor8} & \color{black}{Kirin 950} & \color{black}{\thead{4 $\times$1.80 \\ 4$\times$2.3}} & \color{black}{4} \\ 
		\hline
		\color{black}{\thead{Redmi Note10}} & \color{black}{Snapdragon 678} & \color{black}{\thead{6 $\times$ 1.70   \\  2 $\times$ 2.2}} & \color{black}{6} \\ 
		\hline
		\color{black}{\thead{Google Pixel6}} & \color{black}{Google Tensor} & \color{black}{\thead{4 $\times$ 1.80 \\  2 $\times$ 2.80  \\ 2 $\times$ 2.25 }} & \color{black}{8}\\ 
		\hline
		\color{black}{Oneplus9} & \color{black}{Snapdragon 888} & \color{black}{\thead{6 $\times$ 1.70   \\ 1 $\times$ 2.84 \\ 3 $\times$ 2.42}} & \color{black}{12} \\ 
            \bottomrule[1.5pt]
	\end{tabular}}
	\label{tab:device_hardware} 
\end{table}

\tian{\textbf{Models and Datasets.}
Several well-known DNN and Transformer models in computer vision (CV) and natural language processing (NLP) domains are utilized for evaluation, including 
\begin{itemize}
    \item For image classification:
    \begin{itemize}
        \item Training MNIST~\cite{mnist}, including 60,000 training images and 10,000 test images, handwritten digital images with 10 classes, on LeNet5~\cite{lenet5} model with 60k parameters and AlexNet model with 60M parameters.
        \item Training CIFAR10~\cite{cifar10} consisting of 60000 32x32 color images in 10 classes with 6000 images per class with a training set of 50,000 images and a test set of 10,000 images on VGG16 model with 138M parameters and ResNet18 model with 11.7M parameters.
    \end{itemize}
    \item For text generation:
    \begin{itemize}
        \item Training Shakespeare~\cite{Shakespeare}, a dataset built from The Complete Works of William Shakespeare. It is a popular choice for training language models due to its manageable size and the complexity of Shakespeare's language on an LSTM~\cite{hochreiter1997long} model.
        \item Training Wikitext~\cite{wikitext}, a language modeling dataset that is a collection of over 100 million tokens extracted from the set of verified Good and Featured articles on Wikipedia, on GPT-2~\cite{gpt2}.
    \end{itemize}
\end{itemize}
Dirichlet distribution $p_k \sim Dir_N(\sigma)$ is utilized to simulate the Non-IID data distribution on different devices. It's worth noting that Shakespeare and Wikitext datasets inherently exhibit Non-IID characteristics. We evaluated the performance of GPT-2 using perplexity (PPL), which serves as a metric reflecting the model's text generation capability.}

\begin{figure}
    \centering
    \includegraphics[width = 0.45\linewidth]{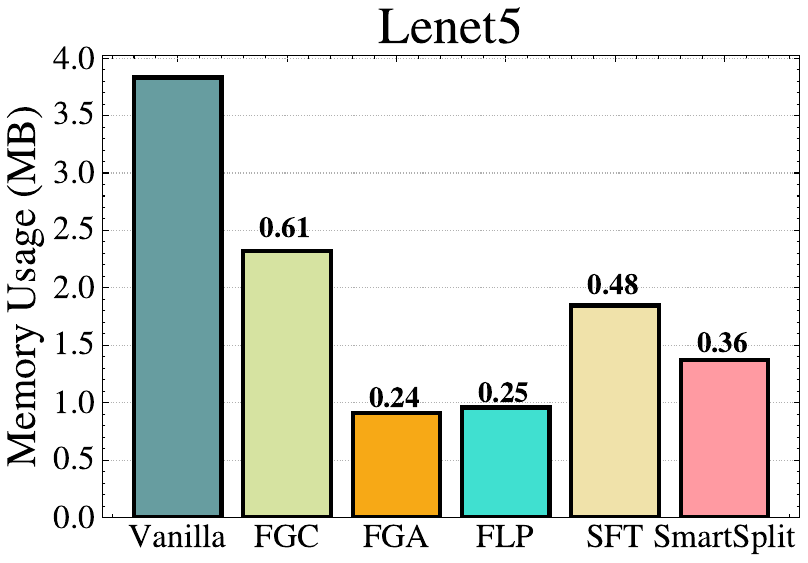}
    \includegraphics[width = 0.45\linewidth]{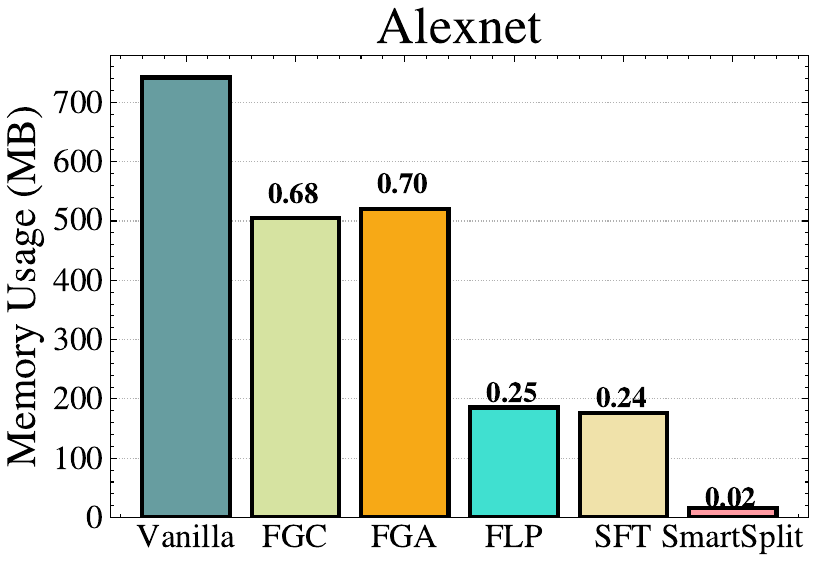}
    \includegraphics[width = 0.45\linewidth]{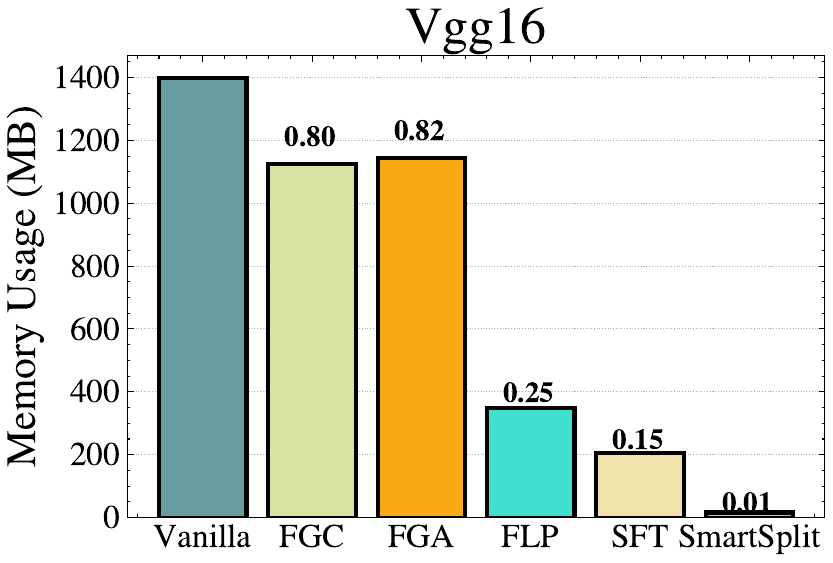}
    \includegraphics[width = 0.45\linewidth]{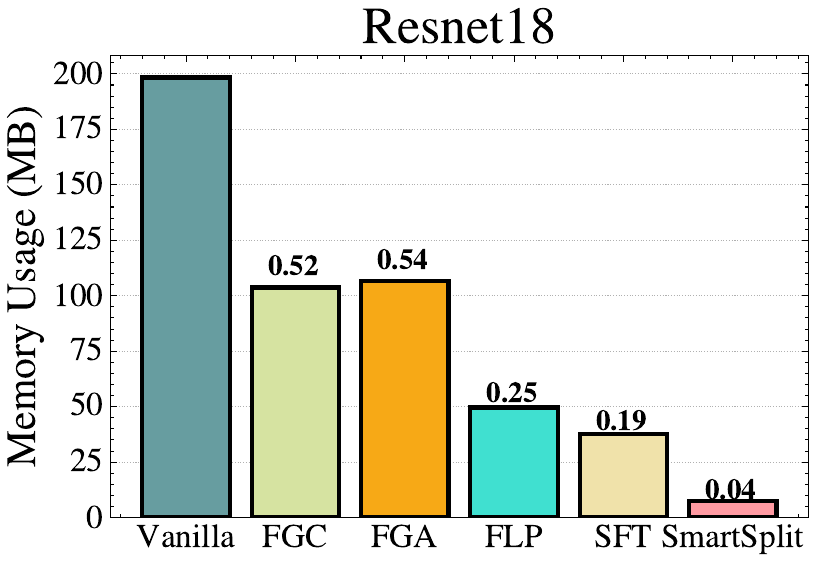}
    \caption{\tian{Minimum memory requirement for FedAvg (Vanilla), gradient checkpointing (FGC), gradient accumulation (FGA),  low precision with int8 (FLP), SplitFL (SFT), and \model~.}}
    \label{fig:evaluation_memory}
\end{figure}

\subsection{Memory Requirements for FL Training}
We initiate our evaluation by detailing the memory requirements for on-device FL training using \model~. We compared and evaluated \model~ with the following \textbf{Group 1} baselines. 
\begin{itemize} 
    \item \textbf{FedAvg}~\cite{fedavg} (vanilla, no memory management): Server randomly selects devices to participate in each training round without considering heterogeneity and memory constraints. 
    \item FL with different memory reduction techniques (Section~\ref{section:memory}), including gradient checkpointing (\textbf{FGC}), gradient accumulation (\textbf{FGA}), and low precision (INT8) (\textbf{FLP}).
    \item SplitFL~\cite{splitfed} (\textbf{SFT}): Merge SL with FL, by leveraging layer offloading from resource-constrained devices to the server for training burden-reducing, without considering the heterogeneous constraints, while adopting a static splitting scheme for all models throughout the whole training process. SplitFL (SFL) statically splits LeNet5 at the second layer (after the 2D MaxPool layer), and second layer of AlexNet, while fourth layer of VGG16, and third layer (after 2D BatchNormalization layer) of ResNet18 as mentioned in~\cite{splitfed}. 
\end{itemize}

\tian{Figure~\ref{fig:evaluation_memory} delineates the memory prerequisites for our four target FL models under various baseline configurations. The data illustrates that \model~ notably diminishes the memory demand for FL model training ( $100\times$ less for VGG16, $50\times$ for AlexNet, $25\times$ for ResNet18, and $2.7\times$ for LeNet5). Model splitting is observed to be more efficacious in cutting down memory usage compared to other memory-saving techniques. With the integration of dynamic memory budget management and cost-aware recomputation, \model~ adeptly overcomes memory limitations, tailoring cross-device FL model training to the constrained memory capacities of mobile platforms.}

\begin{table*}[!t]
    \vspace{-2mm}
  \small
  \centering
  \begin{threeparttable}
  \caption{Model performance of different selection approaches. \model~ achieves the best performance across all the datasets.}
    \textcolor{black}{
    \begin{tabular}{c|cc|cc|c|c}
    \toprule
     \multirow{3}{*}{\begin{minipage}{1cm}
Dataset \\ \&Model
\end{minipage} }  &   \multicolumn{4}{c|}{CV Tasks}    &     \multicolumn{2}{c}{NLP Tasks} \\ \cline{2-7}
& \multicolumn{2}{c|}{MNIST-LeNet5 $\uparrow$} & \multicolumn{2}{c|}{CIFAR10-Resnet18 $\uparrow$} &  \multicolumn{1}{c|}{Shakespeare-LSTM $\uparrow$} & \multicolumn{1}{c}{Wikitext-GPT-2\tnote{1} $\downarrow$}\\ 
      & IID & Non-IID\tnote{2} & IID & Non-IID\tnote{2} & Non-IID\tnote{2} & Non-IID\tnote{2}\\
    \hline  
    FedAvg &  98.84 & 36.70   &   84.78 & 42.02     &  39.18   &   31.13  \\
    TiFL &  98.91 &   91.21  &   84.96  &  56.39 &44.12 &  25.61     \\
    SplitFL & 98.78 & 46.36  & 84.12 & 40.89 & 39.24 &32.09 \\
    FedAdapt  & 98.89 & 87.23 & 85.01 & 50.34 & 44.21 &27.31 \\
    FedVS & 98.82 & 84.34 & 84.35 & 49.12 & 40.21  & 29.81\\
    Oort & $99_{(49)}$\tnote{3}  &    92.47   &   85.69   &  53.51  &   45.42  &    19.12    \\ 
    \hline
    \model~   & \textbf{$99_{(42)}$} & \textbf{93.88} & \textbf{86.27} & \textbf{55.43} & \textbf{46.32} & \textbf{18.71}  \\
    \bottomrule    
    \end{tabular}%
    }
  \label{tab:acc}%
    \begin{tablenotes}
    \scriptsize
   \item[1] \tian{Perplexity (PPL) is utilized to evaluate the GPT-2 model, indicative of its text generation capabilities.}
\item[2] \tian{The coefficient $\sigma$ is set to $0.01$ for MNIST and $0.1$ for other datasets. Shakespeare and Wikitext, which inherently exhibit Non-IID characteristics.}
\item[3] \tian{An early exit strategy is implemented upon reaching the target accuracy (99\%). The numbers in parentheses represent the round at which the exit occurred.}
    \end{tablenotes}
  \end{threeparttable}
\end{table*}%

\subsection{Model Performance}
We compared and evaluated \model~ with the following \textbf{Group 2} baselines on the performance of \model~ in end-to-end training models. 
\begin{itemize}
\item \textbf{FedAvg}: Server \textit{randomly selects} participating devices in each training round without considering heterogeneous environments and resource constraints. 
    \item \textbf{TiFL}~\cite{tifl}: Estimate the \textit{training time} for devices based on the heterogeneous environment and then tier them, while selecting devices from one of the tiers in each training round and tune the tiers' selection probability based on accuracy. 
    \item \textbf{SplitFL}~\cite{splitfed}: Merge SL with FL, by leveraging layer offloading from resource-constrained devices to the server for training burden-reducing, without considering the heterogeneous conditions, while adopting a \textit{static} splitting scheme for all models throughout the whole training process. 
    \item \textbf{FedAdapt}~\cite{fedadapt}: To reduce the training latency of individual participants, the \textit{dynamic} selection and pruning of the training model for the current round based on the observation of the device system configuration utilizing reinforcement learning.
    \item \tian{\textbf{Oort}~\cite{oort}: optimizes device selection by integrating training loss and latency into a user-defined utility. Combines the device \textit{training time} and the importance of the \textit{training samples} as the utility function to determine the subset of devices to train in each round.}
    \item  \tian{\textbf{FedVS}~\cite{li2023fedvs}: trains a model split between the server and the clients with secret sharing schemes for the local data and models.}
\end{itemize}

\tian{For each FL training round, we select 10 devices to participate from a device pool with 100 devices, with $r=50$ training rounds ($r=10$ for GPT-2), and 5 local training epochs per round. And setting Dirichlet distribution to emulate the Non-IID distribution. Set $\sigma =0.01$ for MNIST, $\sigma =0.1$ for others to emulate Non-IID. Shakespeare and Wikitext are naturally Non-IID.}

\tian{
Table~\ref{tab:acc} presents the final test performance of \model~ across all evaluated schemes. The results demonstrate that \model~ consistently outperforms the baseline methods across a variety of domains and tasks. The principle as discussed in Section III, \model~ effective estimation of system and data heterogeneity, selection of optimal participating devices. In particular, for the IID setting, \model~ converges faster and achieves higher accuracy on MINST-LeNet5, improving test accuracy by $1.49\%$ over FedAvg on CIFAR10-ResNet18.
In addition, \model~ is particularly effective on the more challenging Non-IID setting, significantly outperforming the baselines: improving accuracy by $25.91$.
over FedAvg on average, and reducing perplexity by 12.42 (PPL, lower values correspond to stronger text generation capabilities). 
Overall, this experiment demonstrates the practical and robust ability of \model~ to scale to complex workloads and application scenarios.}

\begin{figure}[!ht]
    \centering
    \includegraphics[width = 0.45\linewidth]{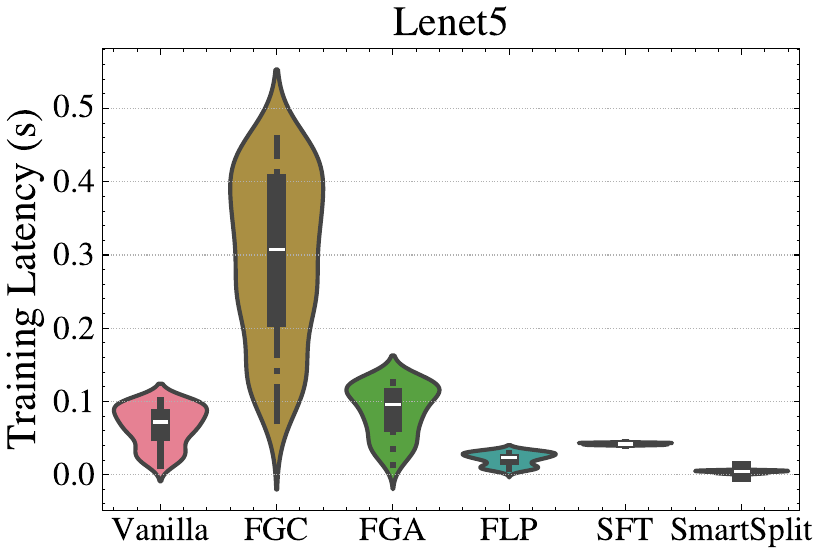}
    \includegraphics[width = 0.45\linewidth]{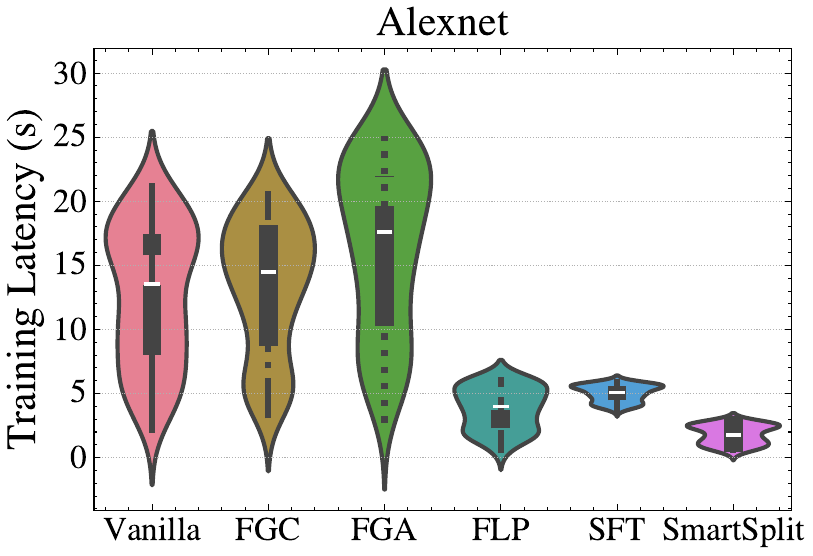}
    \includegraphics[width = 0.45\linewidth]{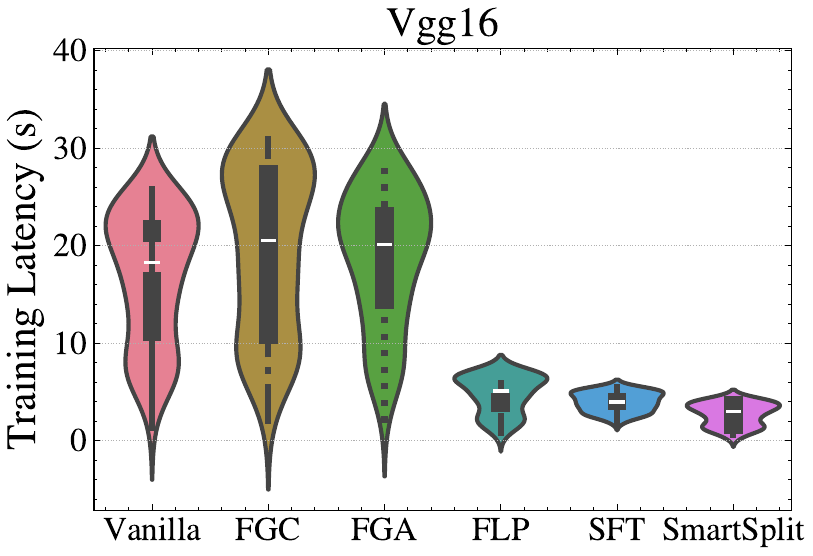}
    \includegraphics[width = 0.45\linewidth]{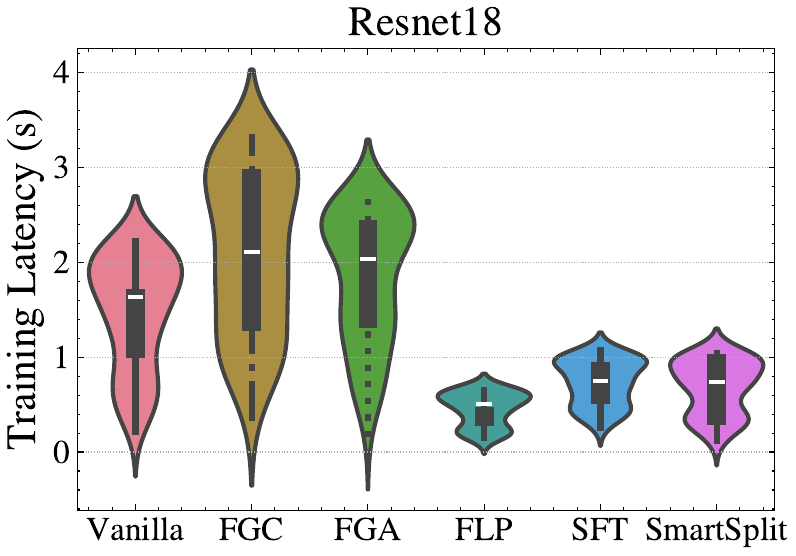}
    \caption{Comparison of per round training time (Group 1).}
    \label{fig:evaluation_time}
\end{figure}
\subsection{Runtime Latency Evaluation}
Utilizing per-round training time as a metric to evaluate the system efficiency, which depends on the computation speed and the scale of required operations executed on the device, as well as the speed of communication IO transmission. Figure~\ref{fig:evaluation_time} represents the per round time overhead during the whole training process. The X-axis represents different frames and Y-axis is the training time. As the figure indicated, \model~ significantly speedup the training process that only needs 6\% training time for the LeNet5 model, 13\% for AlexNet, 16\% for VGG16, and 46\% for ResNet18 compared to the vanilla FL baseline. Doubtlessly, \model~ has high robustness with the lowest training latency and variance throughout the training process due to the consideration of system heterogeneity and time-varying characteristics of statistical importance among devices. 

We can see that different memory reduction techniques introduce other hazards, such as checkpointing and accumulation hampering the convergence speed of the model, and Int8 causing a model performance degradation. SplitFed reduces the computational overhead and memory usage, but frequent communication iterations also block the training efficiency. \model~ is the only one that reduces the training memory usage without degrading the model performance and importing extra overheads that block the training convergence rate.

\subsection{Evaluation of Energy Saving}
\begin{figure}[!t]
    \centering
    \includegraphics[width=0.9 \linewidth]{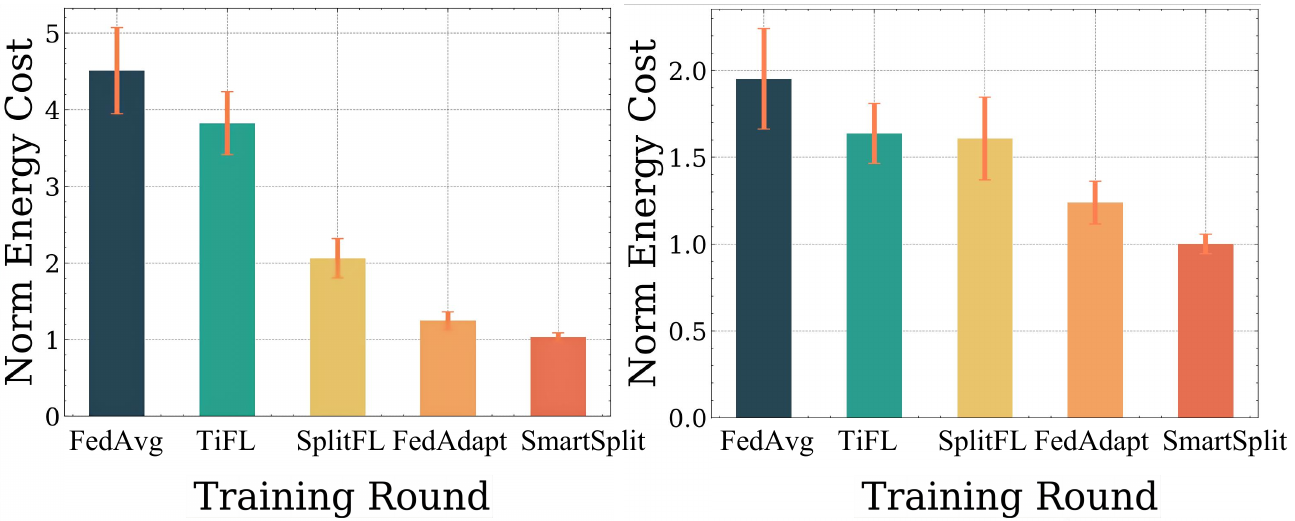}
    \caption{Comparison of the average energy consumption per round. Left: MNIST-LeNet5; Right: CIFAR10-ResNet18.}
    \label{fig:energy}
\end{figure}

\tian{To evaluate the system efficiency on energy cost, a Monsoon Power Monitor is utilized to measure the end-to-end power consumption of the local training process. Figure~\ref{fig:energy} presents the average energy cost and variance across training rounds at different schemes. Obviously, for model-free offloading mechanisms, such as FedAvg and TiFL, their energy consumption is greatly greater than the others. As on MNIST-LeNet5, FedAvg consumes $1.95 \times$ more energy on average than \model~, and $4.51\times$ more on CIFAR10-ResNet18. Meanwhile, since SplitFed with only statically model splitting while without evaluating participating devices for selection, its average energy consumption is $1.65\times$ higher than \model~ and more pronounced at CIFAR10-ResNet18, $2.02\times$ higher. Overall, \model~ shines a light on significant energy savings.} 

\subsection{\tian{Communication Analysis}}
\begin{figure}
    \centering
    \includegraphics[width=0.85\linewidth]{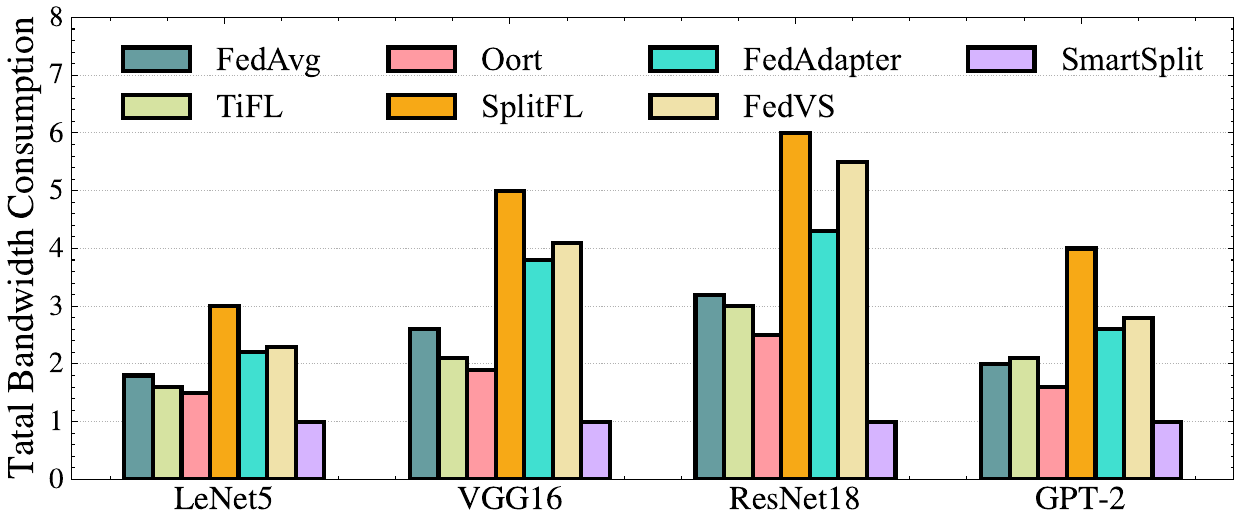}
    \caption{Comparison on normalized communication cost.}
    \label{fig:communication}
\end{figure}

\tian{Then, we evaluate the communication efficiency of \model~. Figure~\ref{fig:communication} illustrates the communication costs for all algorithms, with all values normalized to those of \model~. \textit{Compared to the vanilla split learning architecture}, such as SplitFL, FedVS, and FedAdapter, which use a two-tier split architecture, significant communication costs are incurred due to iterative server-client interactions. In particular, SplitFL and FedVS incur $4.52\times$ and $3.72\times$ the communication costs of \model~ on average, respectively. Despite FedAdapter utilizing dynamic splitting policies to reduce run-time variances, the unstable, low-bandwidth LAN between server and client still hampers the training process. In contrast, \model~ introduces a Device-MEC-Server architecture and a tri-heterogeneity-aware framework, which significantly reduces the communication overhead. \textit{Compared to vanilla federated learning frameworks}, such as FedAvg, TiFL, and Oort, where all model training is performed on mobile devices, these methods still require the entire local model to be sent to the server after each training round, which introduces significant overhead to the local devices. In contrast, SmartSplit trains only partial layers of the model locally, reducing the amount of data to be transmitted and thus minimizing the overall communication cost.}

\subsection{Ablation Studies}
To investigate in-depth a system framework of hierarchical managers, we conduct the following ablation experiments.
1) Static server-device framework as SplitFL (SFT).
2) Dynamic server-device framework considering heterogeneity and runtime variance (D-SFT).
3) Static Server-MEC-device framework (SMD).
4) Static Server-MEC-device framework with importance re-selection (R-SMD).
5) Standard Server-MEC-device framework with no memory management (N-SMD), speed-centric (S-SMD), memory-centric (M-SMD), and cost-aware recomputation (\model~).

\begin{figure}
    \centering
    \includegraphics[width =0.45\linewidth]{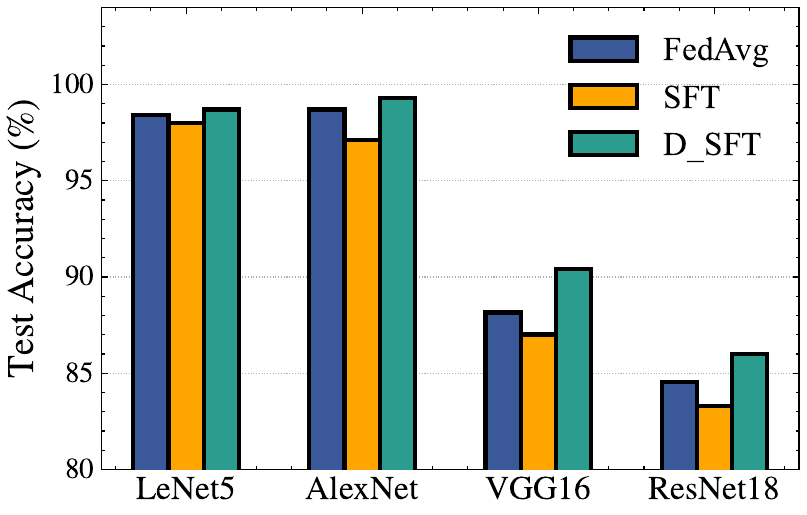}
    \includegraphics[width =0.45\linewidth]{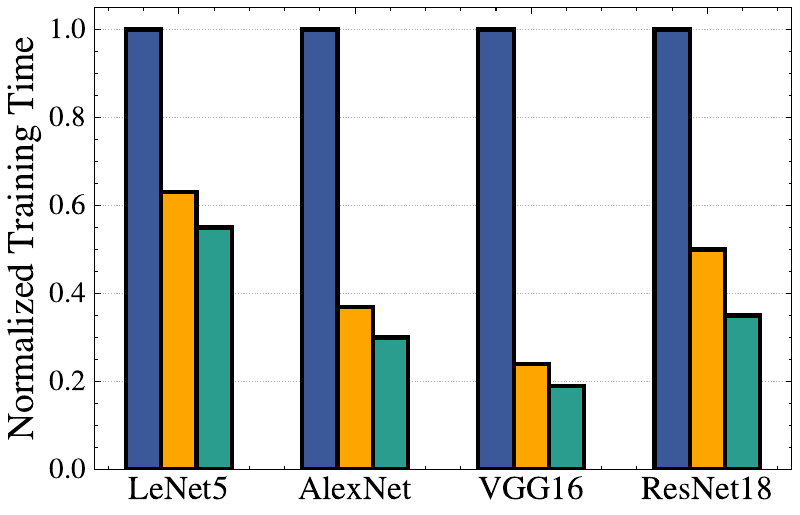}
    \caption{Evaluation of server manager impact on model accuracy and latency. D-SFT: a server-device framework with dynamic splitting.}
    \label{fig:ablation_server}
\end{figure}

\textbf{Central manager.}
Figure~\ref{fig:ablation_server} presents accuracy-to-time results. When comparing Exp.1 and Exp.2, the indispensable role of the central manager within the FL framework becomes evident. By astutely selecting optimal devices and orchestrating layer splitting, it markedly bolsters both system and data efficiency, directly addressing challenge 1 as referenced in section~\ref{challenge}.
Furthermore, model splitting is remarkably effective in reducing training latency, reducing training time to only 24\% to 63\% of that required by fedavg. In addition, the central server introduces significant improvements in model accuracy across different architectures, achieving gains of 1.71\% on LeNet5, 2.12\% on Alexnet, 3.39\% on VGG16, and 2.71\% on Resnet18. These results underscore the critical influence of the central manager and the power of deliberate model partitioning within FL.

\textbf{MEC manager.}
The experimental contrasts between Exp. 1 and 3 underscore the pivotal role of the MEC manager in curbing communication latency, effectively countering challenge 2. Additionally, insights drawn from the evaluation of Exp. 4 indicate that recognizing variances in data importance stands to significantly bolster training velocity. In terms of quantifiable improvements, the MEC manager demonstrated a substantial reduction in training time, ranging from 4\% to 84\%, under conditions facilitated by a high-speed LAN. When further complemented by the implementation of importance re-selection which focuses on filtering out low-contribution samples—the efficiency of the training process is further optimized. Such strategic re-selection resulted in an additional 4-7\% decrease in training time when based on SDM. Collectively, these findings emphasize the efficiency of MEC management and the potential benefits of data importance acknowledgment in expediting the FL training process.
\begin{figure}
    \centering
    \includegraphics[width =0.45\linewidth]{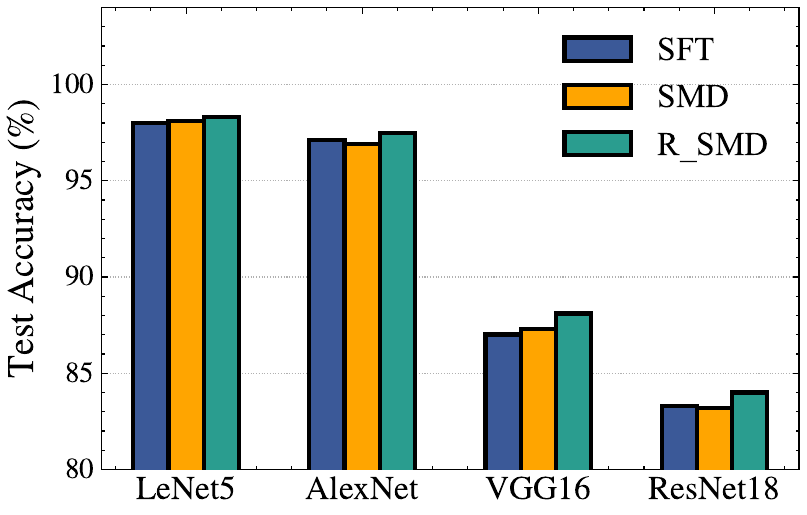}
    \includegraphics[width =0.45\linewidth]{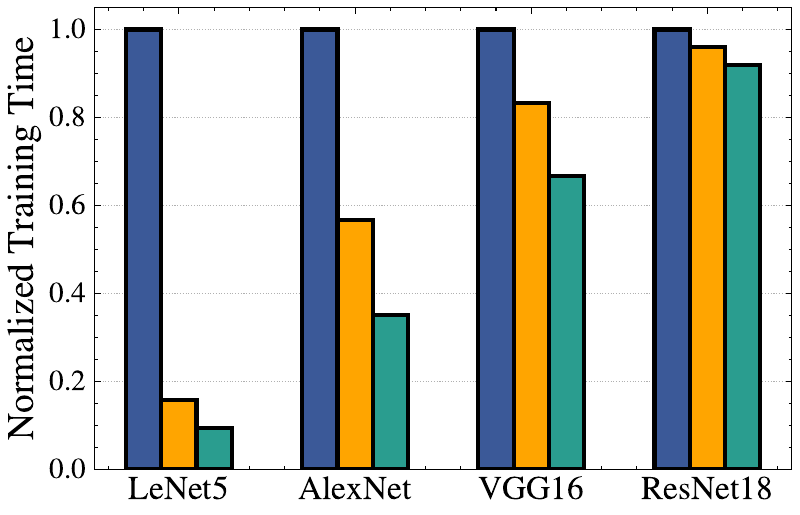}
    \caption{Evaluation of MEC manager impact on model accuracy and latency. SMD: server-MEC-device with static splitting. R-SMD: server-MEC-device with static splitting and importance re-selection.}
    \label{fig:ablation_mec}
\end{figure}

\begin{figure}
    \centering
    \includegraphics[width =0.45\linewidth]{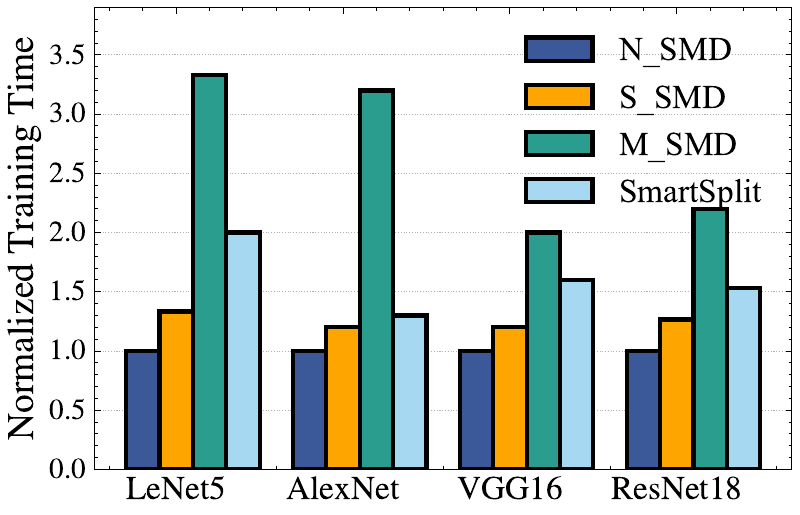}
    \includegraphics[width =0.45\linewidth]{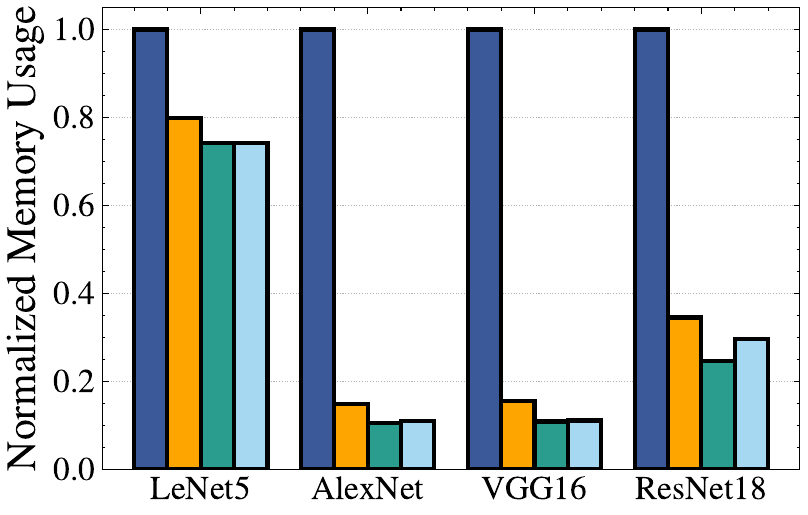}
    \caption{\tian{Evaluation of on-device manager impact on model memory and latency. N-SMD: no memory management. S-SMD: speed-centric recomputation. M-SMD:memory-centric recomputation. \model~: cost-aware recomputation.}}
    \label{fig:ablation_device}
\end{figure}

\textbf{On-device manager.}
This experiment emphasizes on-device management and the advantages of cost-aware dynamic memory budget management, as shown in Figure~\ref{fig:ablation_device}. Cost-aware recomputation combines memory-centric and speed-centric strategies, ensuring $M_{cost} \le l_{peak}$. The memory-centric approach maximizes memory savings but demands more computational resources. The speed-centric approach excels in computational efficiency but may require additional memory. Cost-aware recomputation strikes a balance, maintaining memory savings while reducing the computational burden. Choosing the right strategy is crucial for efficient deep neural network operation on memory-constrained devices, optimizing performance according to specific application needs and available resources.

\subsection{Overhead  Analysis}
	\tian{We provide the overhead analysis for \model~. The primary cost is the time to create training schedules.} Probing training helps acquire device profiles without adding significant computational load. Communication costs for transmitting probing loss values - a mere scalar - are minimal compared to sending full model parameters. \tian{MEC and central managers on edge and central servers don't significantly affect the overall training process.} For mobile training using \model~, VGG16 requires 2.5s and 36.3MB with a batch of 64. Using the Monsoon Power Monitor, VGG16's average consumption is 1.2J per cycle. On a Google Pixel 6, a 10K cycle uses 15\% of the battery, proving \model~ is feasible for mobile on-device FL.

\section{Related Work}
Memory limitations of mobile devices have hindered the practical implementation of FL. Split Learning (SL)~\cite{split_2,split_3,split_4,split_5,split_6} enhances training efficiency and scalability by offloading computation to a server, thus breaking the mobile device wall. \tian{SplitFL~\cite{splitfed} combines FL and SL to minimize cross-device computation and memory overhead. However, existing works~\cite{splitfed, split_2,split_3,split_4,split_5,split_6} primarily rely on ``static'' split positions, ignoring device hardware heterogeneity and runtime environment variance.} Recently, some frameworks~\cite{fedadapt,split_d_2,split_d_3} have explored dynamic splitting, but struggle to address the three core challenges simultaneously. \tian{These challenges include: triple heterogeneity arising from different hardware capabilities, non-IID training data, and different resource demands for model layers; communication bottlenecks in model splitting due to high and frequent communication overhead; and dynamic memory budget issues arising from resource contention during co-running applications.} Such challenges underscore the need for a robust approach to effectively address memory constraints on mobile devices.

\section{Conclusion}
Navigating the challenges of Federated Learning with ever-complex models and device memory wall, this paper introduced \model~. This innovative framework optimizes memory usage without sacrificing training efficiency. Through its hierarchical structure, \model~ dynamically manages device participation, model splitting, and real-time memory budget. Empirical tests validate its effectiveness, demonstrating significant memory savings, reduced latency, and improved model accuracy. Thus, \model~ represents a promising stride towards optimizing FL deployments in memory-constrained environments.

\bibliographystyle{ieeetr}
\bibliography{reference}

\end{document}